\documentclass[12pt]{iopart}

\usepackage{graphicx}
\usepackage{subfigure}
\usepackage{parskip}
\usepackage{float}
\begin{document}

\title[Multi-state Swap Test algorithm]{Multi-state Swap Test Algorithm}

\author{Wen Liu$^{1,2,3}$\& Han-Wen Yin$^2$ \& Zhi-Rao Wang$^2$ \& Wen-Qin Fan$^{1,2}$}

\address{1. State Key Laboratory of Media Convergence and Communication, Communication University of China, Beijing,
China\\
2. School of Computer Science and Cybersecurity , Communication
University of China, Beijing, China \\
3. Key Laboratory of Convergent Media and Intelligent Technology
(Communication University of China), Ministry of Education, Beijing,
China }

\ead{lw$\_$8206@163.com} \vspace{10pt}
\begin{indented}
\item[]April 2022
\end{indented}

\begin{abstract}
Estimating the overlap between two states is an important task with
several applications in quantum information. However, the typical swap test circuit can only measure a sole pair of quantum states at a time.
In this study we designed a recursive quantum circuit to measure overlaps of multiple quantum states $|\phi_1...\phi_n\rangle$concurrently 
with $O(n\log n)$ controlled-swap (CSWAP) gates and $O(\log n)$ ancillary qubits. 
This circuit enables us to get all pairwise overlaps among
input quantum states $|\langle\phi_i|\phi_j\rangle|^2$. Compared
with existing schemes for measuring the overlap of multiple quantum
states, our scheme provides higher precision and less consumption of
ancillary qubits. In addition, we performed simulation experiments on
IBM quantum cloud platform to verify the superiority of the scheme.
\end{abstract}

\vspace{2pc} \noindent{\it Keywords}:Quantum information, The Swap Test, Multiple Quantum
States, Overlap

\submitto{\JPA}
\maketitle

\section{Introduction}

The concept of quantum computing was proposed by Richard Feynman in the early 1980s\cite{1} and David Deutsch later gave a quantum algorithm solution to a toy issue\cite{2}.
In recent years, Quantum algorithms and their applications in fields like encryption\cite{3},database search\cite{4}, quantum simulation\cite{5}, optimization
problems\cite{6} and linear systems of equations\cite{7} (refer to
\cite{8} for more information) have proved spectacular.
For certain problems, these quantum algorithms outperform classical algorithms in terms of speed.
Typically, quantum algorithms are described using quantum circuits, which are made up of a sequence of quantum gates that can handle quantum bits.
When dealing with quantum circuits, it's important to measure the similarity or overlap between two states $\left| \phi \right\rangle$ and $\left| \psi \right\rangle$, which can be denoted by $\left| {\left\langle {\phi ,\psi} \right\rangle } \right|^2$\cite{9}.

Swap Test is a well-known quantum circuit, which uses a controlled-swap (CSWAP) gate and an ancillary qubit to estimate the inner product $\left| {\left\langle {\phi ,\psi } \right\rangle } \right|^2$ of two states $\left|\phi \right\rangle ,\left| \psi \right\rangle$\cite{10}\cite{11}. It has attracted interest as a fundamental primitive and its efficient implementation on near-term quantum computers is a hot research issue at the moment. In\cite{12}, L. Cincio et al. presented a machine learning approach
for discovering short-depth algorithm to compute states overlap, whose performance is better than the linear scaling of Swap Test. In \cite{13}, M. Fanizza et al. studied the estimation of overlap between two unknown pure quantum states in a finite-dimensional system, given $M$ and $N$ copies of each type. They demonstrated that allowing for general collective measurements on all copies can yield a more precise estimate. In \cite{14}, U. Chabaud et al. proposed a strategy for a programmable projective measurement device, which may be interpreted as an optimal Swap Test when only one copy of one state and $M-1$ of the other is available.
There are many applications that require state overlap computation thus making Swap Test appears as a subroutine in these applications. For instance, it is
used to measure entanglement and indistinguishabiliy for quantum
states\cite{11}\cite{15}, and also common in quantum supervised
learning to measure the distances of quantum
states\cite{16}\cite{17}\cite{18}. What's more, proofs-of-concept in
implementations of quantum computers\cite{19}\cite{20} call for this Swap Test as well.

Consider the case when there are $n$ quantum states$\left| {\phi
_1 }\right\rangle ,\left| {\phi _2 }\right\rangle
,...,\left|{\phi _n } \right\rangle$ and the Swap Test can only
assess the overlap of two at a time. In order to obtain the overlap
$\left| {\left\langle {\phi _i } \right|\left. {\phi _j
}\right\rangle } \right|^2$ between arbitrary two states
$\left|{\phi _i } \right\rangle ,\left| {\phi _j
}\right\rangle(i,j=1,2,...,m;i \ne j)$, there is a trivial solution
which requires $\frac{{n(n - 1)}}{2}$ Swap Tests and $\frac{{n(n -
1)}}{2}$ ancillary qubits. In \cite{21}, X. Gitiaux et al. extended
two-state Swap Test to $n$ quantum states $\left|{\phi _1 }
\right\rangle ,\left| {\phi _2 } \right\rangle,...,\left|
{\phi _n } \right\rangle$. They built a unitary ${\rm U}_4$ circuit
with three ancillaries, three controlled-swap (CSWAP) gates, and one
Swap Test to compute overlap for four states. Based on the ${\rm U}_4$
circuit, they developed a recursive algorithm to generalize a
circuit which can estimate overlaps between $n$ quantum states at a
time. The circuit requires $O(n)$ CSWAP gates and $O(\log n)$
ancillary qubits.

In this paper, following the idea of \cite{21}, we design a new
circuit ${\rm U}_4$ with two ancillaries, two controlled-swap(CSWAP) gates
, and two simple Swap Test. We also present two rules to swap four
groups of quantum states. Based on the swap rules and our ${\rm U}_4$, a
circuit to calculate the overlap between different pairs of $n$
arbitrary input states are presented. Our circuit requires $n\log_2n$
CSWAP gates and $2(n-1)$ ancillary qubits.

The paper is organized as follows: some correlative preliminaries are introduced in Section 2; a multi-state Swap Test algorithm is proposed in Section 3; In Section 4, the simulation of 8-state Swap Test algorithm on IBM Quantum Experience's platform is described. And the correctness and performance of the algorithm are analyzed in section 5. A brief discussion and a summary are given in Section 6.

\section{Preliminaries}
  \subsection{Swap Test and Its Quantum Circuit}

The Swap Test is used to estimate the overlap between two unknown quantum states, and it has a 
variety of applications. Given two unknown quantum states $|\phi\rangle$ and $|\psi\rangle$, 
the overlap between them is defined as $|\langle\phi|\psi\rangle|^2$, i.e. the inner product 
between their state vectors. The Swap Test quantum circuit for estimating the overlap between 
$|\phi\rangle$ and $|\psi\rangle$ is shown in Fig.(1):

  \begin{figure}[htbp]
    \includegraphics[height=2.6cm,width=6cm]{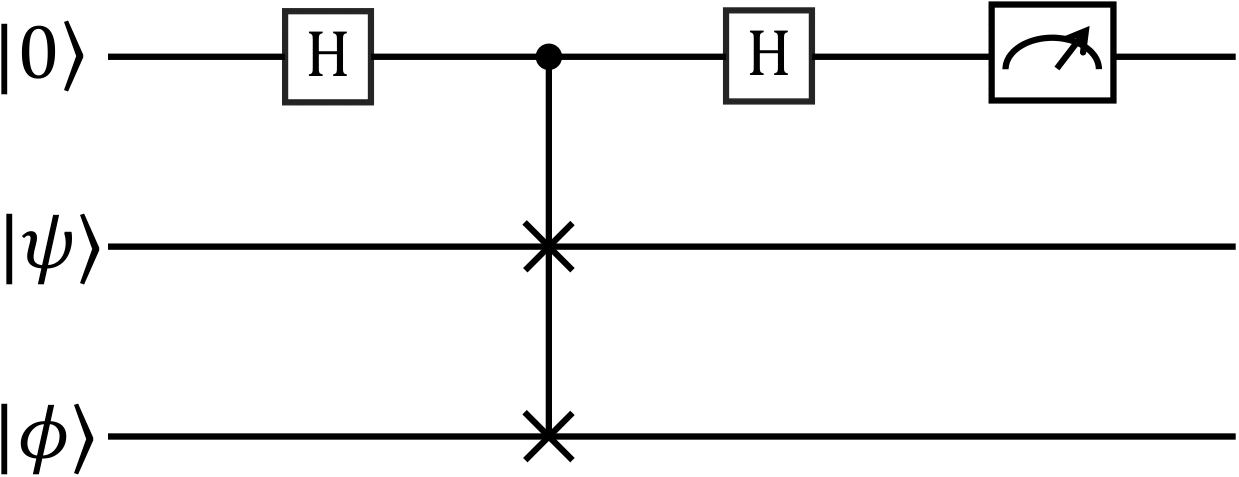}
    \centering
    \caption{Quantum circuit implementing the Swap Test for two quantum states}
  \end{figure}

where the first qubit is an auxiliary qubit used for measurement and the other two qubits 
are used for input. According to the quantum circuit, the output denoted by $|\rho\rangle$ can be calculated as:

\begin{equation}
\begin{array}{l}
    |\rho\rangle=|0\rangle|\phi\rangle|\psi\rangle\\
    \rightarrow {1\over\sqrt{2}}(|0\rangle+|1\rangle)|\phi\rangle|\psi\rangle\\
  \rightarrow{1\over\sqrt{2}}(|0\rangle|\phi\rangle|\psi\rangle+|1\rangle|\psi\rangle|\phi\rangle)\\
   \rightarrow{1\over2}|0\rangle\left(|\phi\rangle|\psi\rangle+|\psi\rangle|\phi\rangle\right)+{1\over2}|1\rangle\left(|\phi\rangle|\psi\rangle-|\psi\rangle|\phi\rangle\right)
 \end{array}
    \end{equation}

According to Equ.(1), the probability of measuring 
$|0\rangle$ is $
Prob(0) = \frac{{1 + | {\langle \phi  | \psi \rangle } |^2
}}{2}$.

Obviously, the overlap $\langle\phi|\psi\rangle$ is related to the probability of measuring $|0\rangle$. So we can derive:

\begin{centering}
  \begin{equation}
    |\langle\phi|\psi\rangle|^2=2Prob(0)-1
  \end{equation}
\end{centering}

In \cite{15}, Juan Carlos Garcia-Escartin et al. show that the Hong-Ou-Mandel effect from quantum optics is equivalent to the Swap Test. Then, they give a destructive Swap Test module that can provide better performance than the traditional Swap Test.
The basic implementation of their Swap Test costs three quantum registers and a CSWAP gate to estimate the overlap between two unknown quantum states. The CSWAP circuit is inspired by classical XOR swapping, and its quantum implementation is shown in Fig.(2).

\begin{figure}[htbp]
      \includegraphics[height=2.8cm,width=6cm]{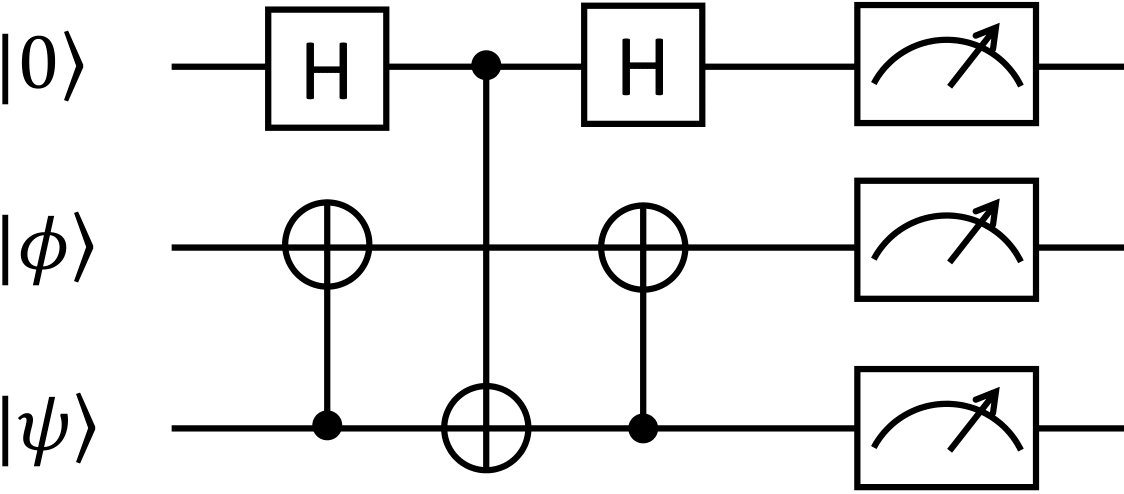}
        \centering
  \caption{Swap Test for qubit states with a CSWAP gate.}
\end{figure}

X gate can get with the help of Z gate.
So we have the equivalent quantum circuit of Swap Test in Fig.(3).

\begin{figure}[htbp]
    \includegraphics[height=2.8cm,width=6.5cm]{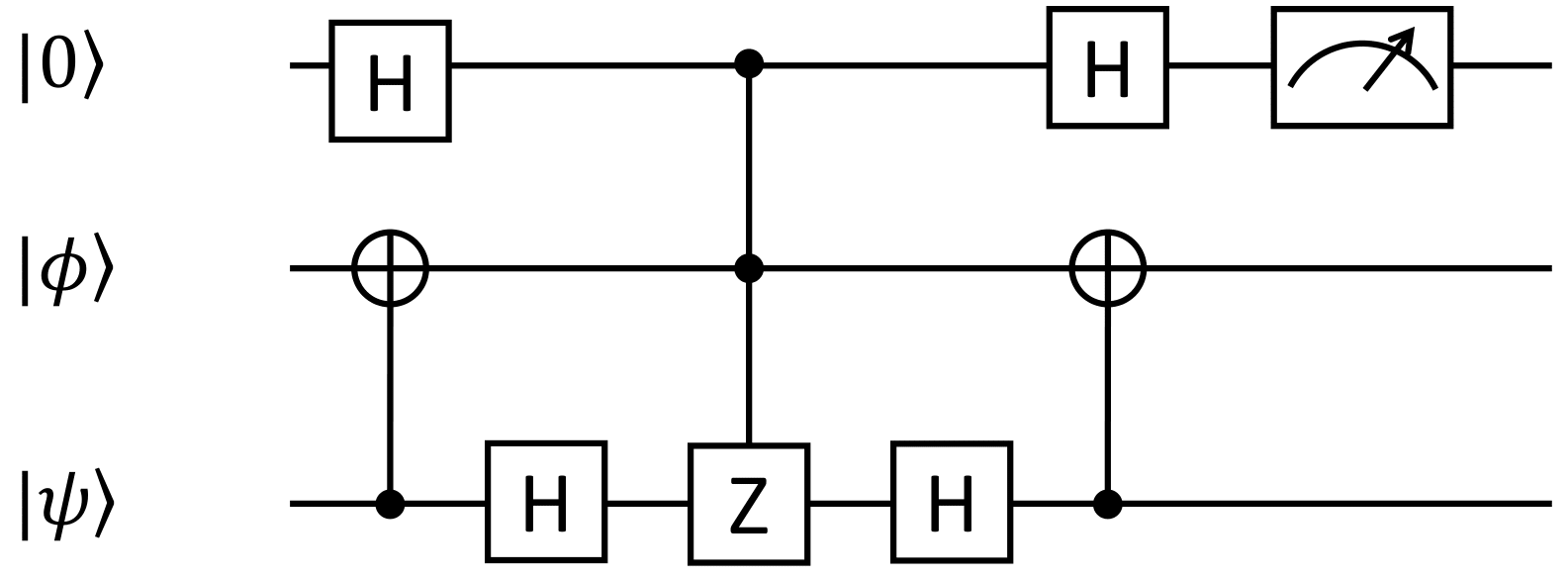}
    \centering
\caption{Swap Test circuit with a CCZ gate.}
\end{figure}

Obviously, the ancillary qubit is not affected by the tested qubits after CCZ gate. We can just get rid of the last H and CNOT gates and perform measurements directly after the CCZ gate with no effect on the ancillary qubit and the result of the Swap Test. Moreover, with the help of a trick called
'phase kickback', we can rewrite the circuit as in Fig.(4).

\begin{figure}[htbp]
    \includegraphics[height=2.8cm,width=5cm]{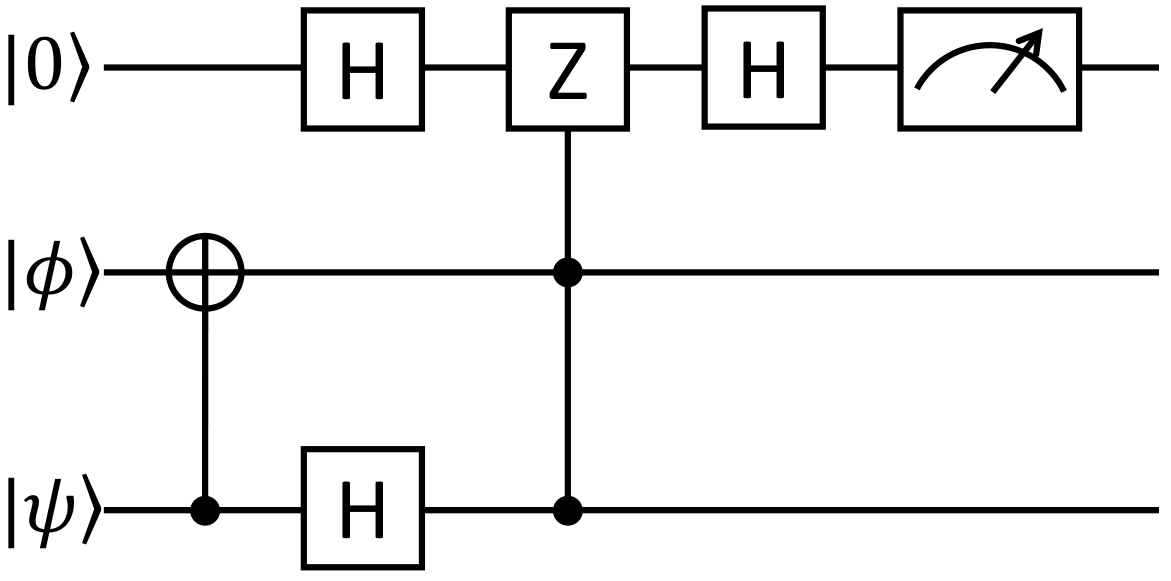}
      \centering
\caption{Swap Test advancing the measurement}
\end{figure}

The output of ancillary qubit can be predicted according to the output of tested qubits in the classical method, which is sometimes called the principle of deferred measurement. Clearly, we can just ignore the ancillary qubit and perform a Swap Test with the circuit in Fig.(5), which is the improved Swap Test.

\begin{figure}[htbp]
    \includegraphics[height=2cm,width=4cm]{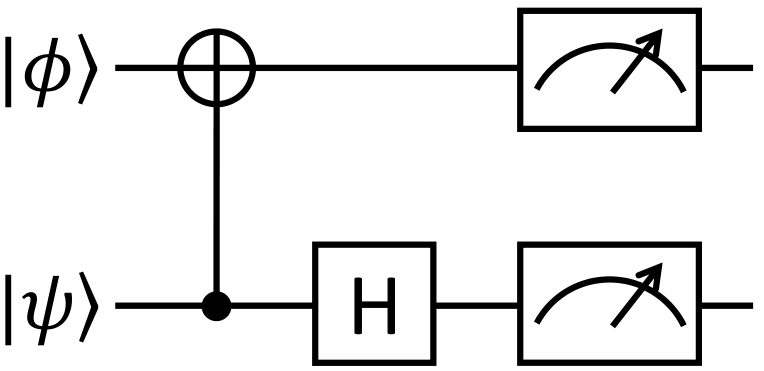}
    \centering
\caption{Destructive Swap Test}
\end{figure}

\subsection{Gitiaux's Swap Test for Multiple Quantum States}

In \cite{19},  Xavier Gitiaux et al. gave a new quantum algorithm that is used to estimate the overlap between multiple quantum states based on the standard two-state Swap Test and CSWAP gate.

First, they define a unitary ${\rm U}_4$ for a 4-state circuit as a building
block to develop a recursive algorithm to produce an $n$-state circuit.

\begin{figure}[htbp]
    \includegraphics[height=6cm,width=4.5cm]{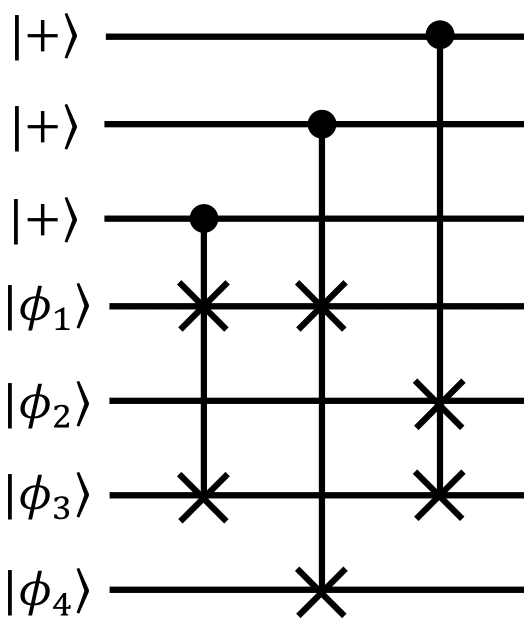}
      \centering
\caption{Quantum circuit implementing ${\rm U}_4$ for 4 quantum states }
\end{figure}

As shown in Fig.(6), there are a total of four input states which can form 6 pairs of combinations.
It requires at least 6 orthogonal states to label them. Therefore, the minimum number of ancillary qubits is three, and the minimum number of CSWAP gates is also three, each acting as control once.

Terms contained in the superposition state are listed in Table 1, where $s_{1},s_{2},s_{3}$ denote the 
basis states of the three ancillary qubits from bottom to top in the ancillary section in Fig.(6). The 
four state registers are labeled by $q_i(i=1,2,3,4)$. The input state is $|\phi_1\phi_2\phi_3\phi_4\rangle$, and the order of the output states depends on the measured values of the auxiliary qubits. As shown in the table 1, the output state will be $|\phi_4\phi_3\phi_2\phi_1\rangle$ if the ancillary qubits are $|011\rangle$.The essential of this algorithm is that each possible overlap between two states can be generated by only
measuring the first two qubits, which can be verified by Table.(1).

\begin{table}[htbp]
\centering
\begin{tabular}{|l|l|l|l|l|}
\hline
$s1s2$ & $|00\rangle$ & $|01\rangle$ & $|10\rangle$ & $|11\rangle$ \\ \hline
$s3=|0\rangle$ & $|\phi_1\phi_2\phi_3\phi_4\rangle$ & $|\phi_1\phi_4\phi_3\phi_2\rangle$ & $|\phi_3\phi_2\phi_1\phi_4\rangle$ & $|\phi_4\phi_2\phi_1\phi_3\rangle$ \\ \hline
$s3=|1\rangle$ & $|\phi_1\phi_3\phi_2\phi_4\rangle$ & $|\phi_4\phi_3\phi_2\phi_1\rangle$ & $|\phi_3\phi_1\phi_2\phi_4\rangle$ & $|\phi_4\phi_1\phi_2\phi_3\rangle$ \\ \hline
\end{tabular}
\caption{The swap results of $s_1,s_2,s_3$ and $q_1,q_2,q_3,q_4$}
\end{table}

\begin{figure}[htbp]
\includegraphics[height=5cm,width=5.5cm]{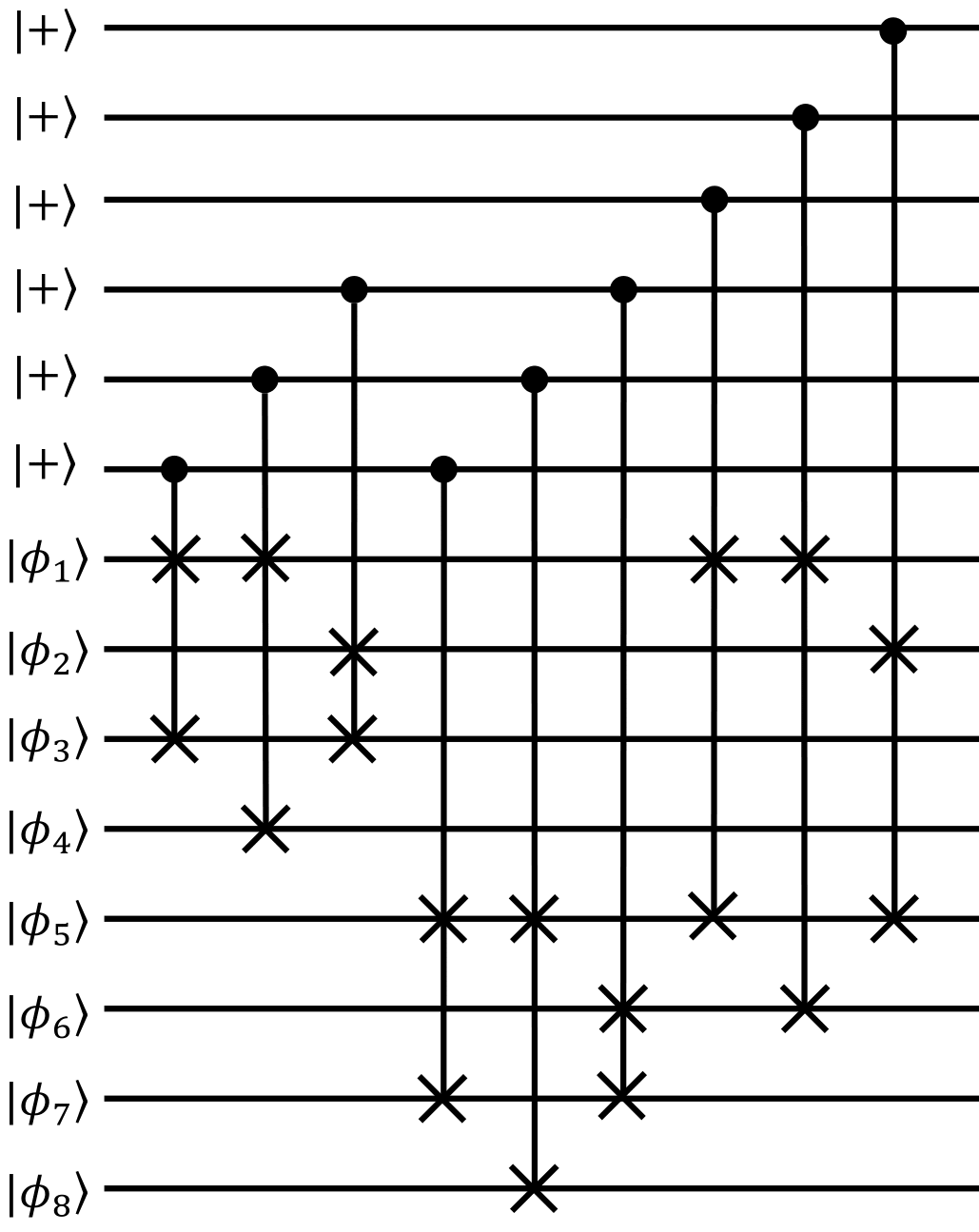}
\centering
\caption{Quantum circuit implementing ${\rm U}_8$ for 8 quantum states}
\end{figure}
And the ${\rm U}_8$ for 8 inputs can be constructed from ${\rm U}_4$ which is shown in Fig.(7). The circuit is
built into two sections of three ancillaries and three groups of three CSWAP
gates, which amounts to six ancillaries and nine CSWAP gates. It is obvious
that the quantum circuit is divided into two groups of four registers: $q_1q_2q_3q_4$
and $q_5q_6q_7q_8$ respectively. In the first step, the corresponding registers in the two 
sets are controlled by a set of three ancillaries in parallel which brings all six pairs in
$|\phi_1\rangle,|\phi_2\rangle,|\phi_3\rangle,|\phi_4\rangle$ to $q_1q_2$ and all six pairs in $|\phi_5\rangle,|\phi_6\rangle,|\phi_7\rangle,|\phi_8\rangle$ to $q_5q_6$ respectively.
Then another set of three ancillaries is introduced to control the pairing among $q_1q_2q_5q_6$ which
makes the pairing between all 8 registers possible.

Next we consider the case $n=2^k$. As shown in Fig.(8), we build the
${\rm U}_n$ by repeating the same thing we did in ${\rm U}_8$. Denoting the
number of ancillaries in ${\rm U}_n$ as $d_n$ and the number of CSWAP gate
as $c_n$, We have $d_n=3k-3$ and $c_n=3(2^{k-1}-1)$ where
$k=log_2n$.

\begin{figure}[htbp]
    \includegraphics[height=5cm,width=5.5cm]{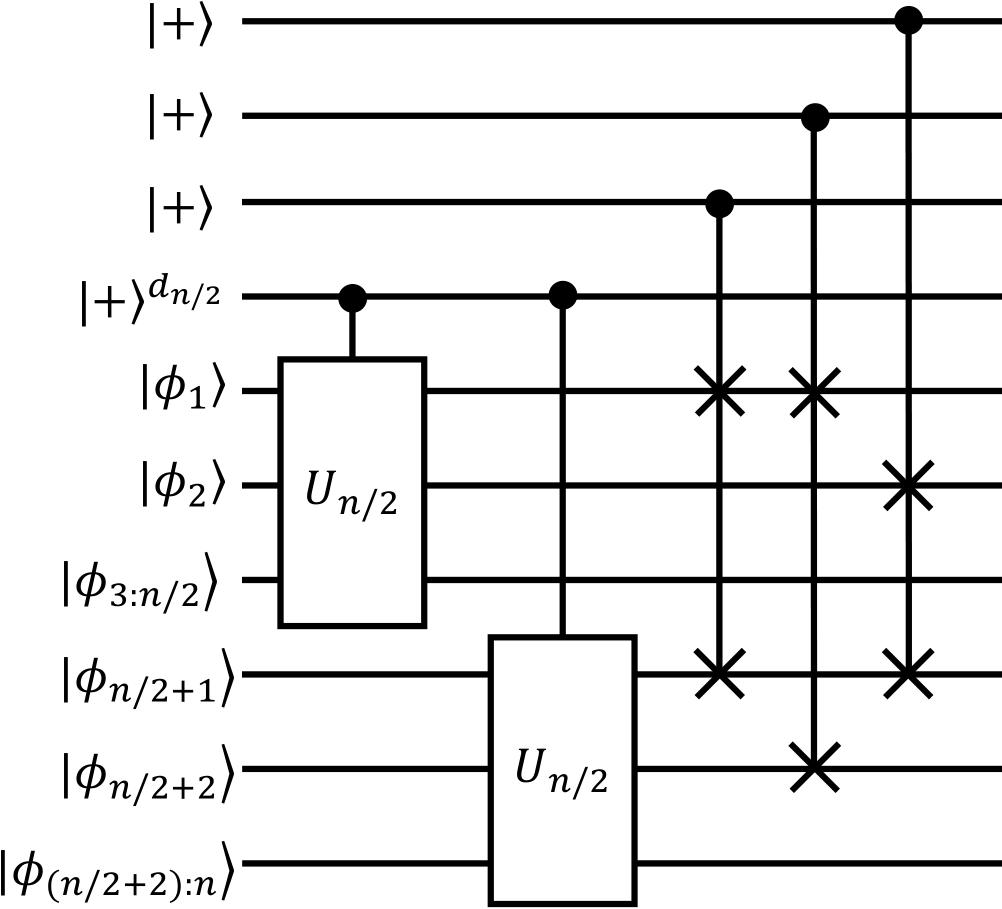}
      \centering
\caption{Quantum circuit implementing ${\rm U}_n$ for n quantum states}
\end{figure}

To clarify, the procedure is to use ${\rm U}_4$ within different groups first, and then on the resultant registers of each group. ${\rm U}_4$ acts each time to allow the pairing of two parts of the registers(these parts could be the simplest group or a group already combined by ${\rm U}_4$ in the previous step.)

\section{New Multi-state Swap Test algorithm}

Suppose that there are $n$ quantum states $| {\phi_1 }
\rangle ,| {\phi_2 }\rangle ,...,|{\phi_n }\rangle$. An algorithm to construct one circuit
to take all $n$ input states at once and get the overlap of arbitrary two input states in $| {\phi _1 }\rangle ,| {\phi _2 } \rangle ,...,|{\phi_n }\rangle$ is presented.

Two swap rules are designed as follows: $N(N=2^2,2^3,...,2^k)$
quantum states can be divided into four groups $G_1,G_2,G_3,G_4$.
$Rule 1$ is to exchange the states of 2nd and 3rd group in turn and
get $G_1,G_3,G_2,G_4$; $Rule 2$ is to exchange the states of 2nd and
4th group in turn and get $G_1,G_4,G_3,G_2$. The multi-state Swap
Test algorithm based on the two swap rules is as follows:

(1) Constructing a circuit ${\rm U}_4$ to estimate the overlap of any two of the 4 states quantum states $| {\phi _1 } \rangle ,|{\phi _2 }\rangle ,| {\phi_3 }\rangle,| {\phi _4 }\rangle$.  In ${\rm U}_4$, there are two ancillary qubits $s_1,s_2$ and four input qubits $q_1,q_2,q_3,q_4$.
The initial states are $s_1s_2=|+\rangle|+\rangle$ and $q_1q_2q_3q_4=| {\phi_1 }\rangle
| {\phi _2 } \rangle| {\phi _3 }\rangle | {\phi_4 }\rangle$. $q_1q_2q_3q_4$ are divided into four groups $G_1=q_1;G_2=q_2;G_3=q_3;G_4=q_4$. There are also two CSWAP gates, where $s_1(s_2)$ is the control qubit to swap two groups $G_2,G_3(q_2,q_4)$ according to $rule
1(rule 2)$. Two simple Swap Test for two quantum states are places at the end of ${\rm U}_4$ to measure the overlap for $q_1,q_2$ and $q_3,q_4$. The ${\rm U}_4$ is shown in Fig.(9).
\begin{figure}[htbp]
\includegraphics[height=3.6cm,width=4.2cm]{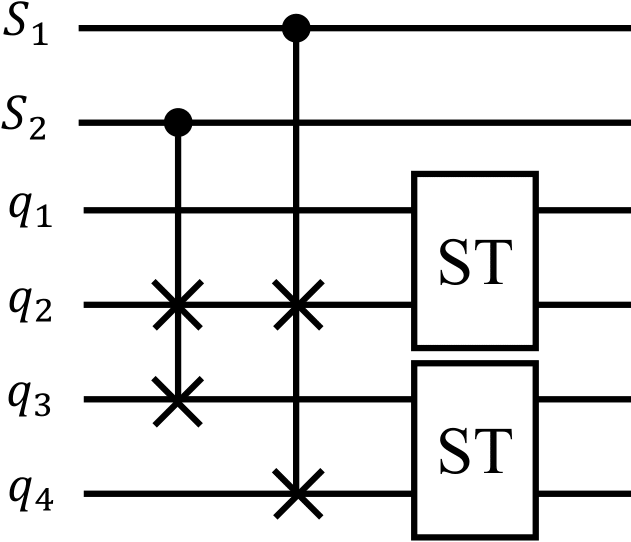}
\centering
\caption{Quantum circuit implementing ${\rm U}_4$ for 4 quantum states}
\end{figure}

The swap results between $s_1,s_2$ and $q_1,q_2,q_3,q_4$ are shown
in Table.(2).
\begin{table}[htbp]\normalsize
\centering
  \begin{tabular}{|l|l|l|l|l|}
  \hline
  $s_1s_2$ & $\left| {0}  {0} \right\rangle$ & $\left| {0}  {1}
\right\rangle$   & $\left| {1}  {0}
\right\rangle$ & $\left| {1}  {1} \right\rangle$
\\ \hline
  $q_1q_2q_3q_4$ & $\left| {\phi _1 \phi _2 }
 {\phi _3 \phi _4 } \right\rangle $ & $
\left| {\phi _1 \phi _3 }  {\phi _2
\phi _4 } \right\rangle$ & $\left| {\phi _1 \phi _4 }
 {\phi _3 \phi _2 } \right\rangle $ & $
\left| {\phi _1 \phi _4 }  {\phi _2
\phi _3 }
\right\rangle$ \\
\hline
  \end{tabular}
  \caption{The swap results of $s_1,s_2$ and $q_1,q_2,q_3,q_4$}
\end{table}

According to the Table.(2), if $s_1s_2=\left| {0}
 {0} \right\rangle$, the overlap of $\left|
{\phi _1 \phi _2 } \right\rangle $ and $\left| {\phi _3
\phi _4 } \right\rangle $ can be obtained through two simple Swap
Test; if $s_1s_2=\left| {0}  {1} \right\rangle$,
the overlap of $\left| {\phi _1 \phi _3 } \right\rangle$ and
$\left| {\phi _2 \phi _4 } \right\rangle $ can be obtained
through two simple Swap Test; if $s_1s_2=\left| {1}
 {0} \right\rangle$, the overlap of $\left|
{\phi _1 \phi _4 } \right\rangle $ and $ \left| {\phi _3
\phi _2 } \right\rangle $ can be obtained through two simple Swap
Test.

(2) Constructing a circuit ${\rm U}_8$ to estimate the overlap of any two quantum states in $\left| {\phi _1 } \right\rangle ,\left|
{\phi _2 } \right\rangle ,\left| {\phi _3 } \right\rangle
,\left| {\phi _4 } \right\rangle,\left| {\phi _5 }
\right\rangle ,\left| {\phi _6 } \right\rangle ,\left| {\phi
_7 } \right\rangle ,\left| {\phi _8 } \right\rangle$. In ${\rm U}_8$,
there are four ancillary qubits $s_1,s_2,s_3,s_4$ and eight input
qubits $q_1,q_2,q_3,q_4,q_5,q_6,q_7,q_8$. The initial states are
$s_1s_2s_3s_4=\left|  + \right\rangle\left|  + \right\rangle\left| +
\right\rangle\left|  + \right\rangle$ and
$q_1q_2q_3q_4q_5q_6q_7q_8=\left| {\phi _1 } 
{\phi _2 }  {\phi _3 } 
{\phi _4 }  {\phi _5 } 
{\phi _6 }  {\phi _7 } 
{\phi _8 } \right\rangle$. $q_1q_2q_3q_4q_5q_6q_7q_8$ are divided
into four groups $G_1=q_1q_2;G_2=q_3q_4;G_3=q_5q_6;G_4=q_7q_8$.
There are also eight CSWAP gates. $s_1(s_2)$ is a control qubit on
two CSWAP gates to swap $G_2,G_3(G_2,G_4)$ according to $rule
1(rule 2)$. And $s_3,q_1,q_2,q_3,q_4$ can be seen as inputs of
${\rm U}_4$ and $s_4,q_5,q_6,q_7,q_8$ can also be seen as inputs of ${\rm U}_4$.
Four simple Swap Test for two quantum states are places at the end
of ${\rm U}_8$ to measure the overlap for $q_1,q_2$ and $q_3,q_4$. The
${\rm U}_8$ is shown in Fig.(10).

\begin{figure}[htbp]
\includegraphics[height=5cm,width=5.5cm]{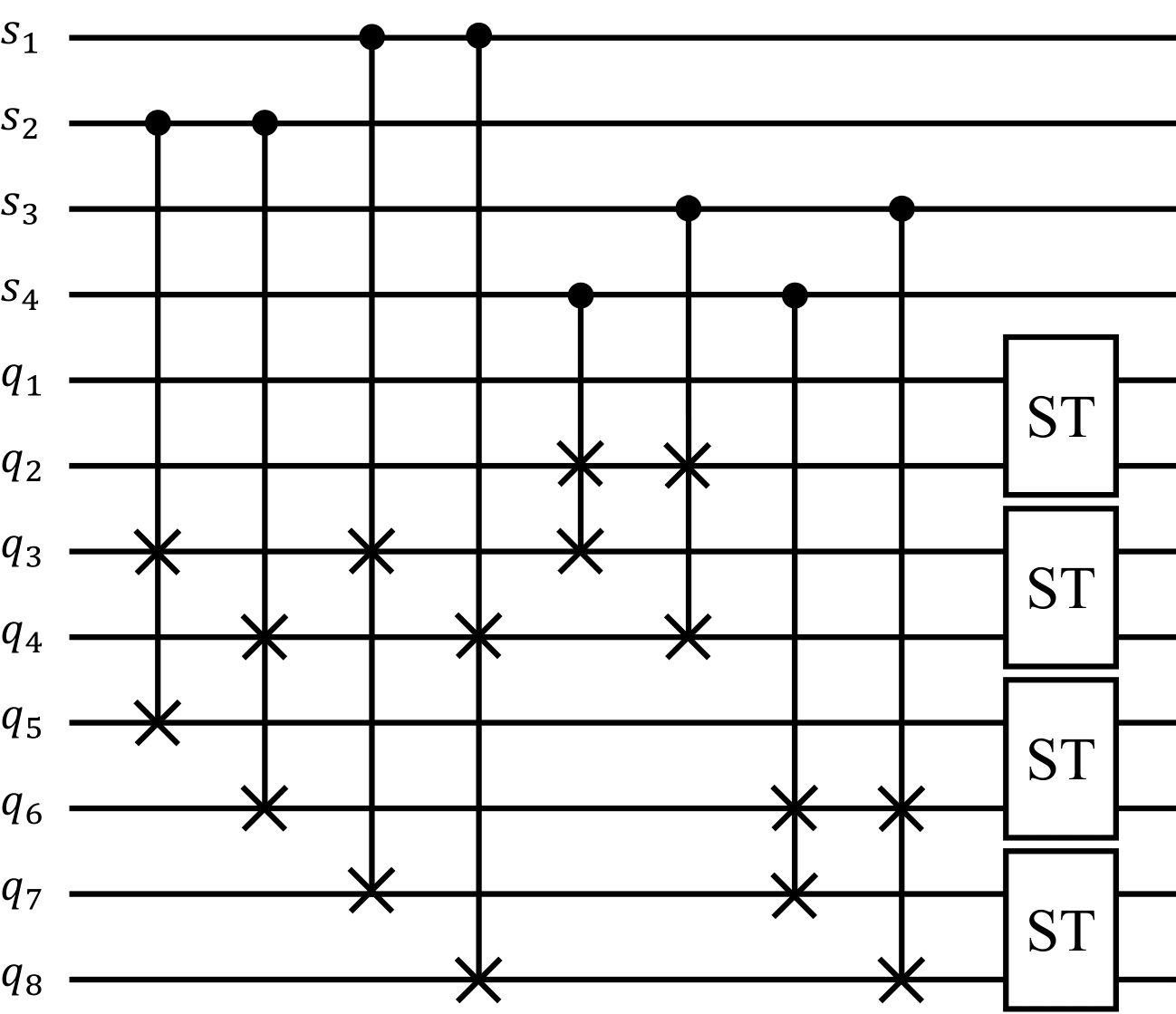}
\centering
\caption{Quantum circuit implementing ${\rm U}_8$ for 8 quantum states}
\end{figure}

The swap results between $s_1,s_2,s_3,s_4$ and
$q_1,q_2,q_3,q_4,q_5,q_6,q_7,q_8$ are shown in Table.(3).
\begin{table}[htbp]
\resizebox{16cm}{1.8cm}{
  \begin{tabular}{|l|l|l|l|l|}
  \hline
  $s_1s_2s_3s_4$ & $\left| {0000} \right\rangle $& $\left| {0001} \right\rangle $  & $\left| {0010} \right\rangle $ & $\left| {0011} \right\rangle $
\\ \hline
  $q_1q_2q_3q_4q_5q_6q_7q_8$ &$
\left| {\phi _1 \phi _2 } \right\rangle \left| {\phi _3
\phi _4 } \right\rangle \left| {\phi _5 \phi _6 }
\right\rangle \left| {\phi _7 \phi _8 } \right\rangle$ &
$\left| {\phi _1 \phi _3 } \right\rangle \left| {\phi _2
\phi _4 } \right\rangle \left| {\phi _5 \phi _7 }
\right\rangle \left| {\phi _6 \phi _8 } \right\rangle  $ &
$\left| {\phi _1 \phi _4 } \right\rangle \left| {\phi _3
\phi _2 } \right\rangle \left| {\phi _5 \phi _8 }
\right\rangle \left| {\phi _7 \phi _6 } \right\rangle  $&
$\left| {\phi _1 \phi _4 } \right\rangle \left| {\phi _3
\phi _2 } \right\rangle \left| {\phi _5 \phi _8 }
\right\rangle \left| {\phi _7 \phi _6 } \right\rangle  $\\
\hline
  $s_1s_2s_3s_4$ & $\left| {0100} \right\rangle $        & $\left| {0101} \right\rangle $  & $\left| {0110} \right\rangle $ & $\left| {0111} \right\rangle $
\\ \hline
  $q_1q_2q_3q_4q_5q_6q_7q_8$ &$
\left| {\phi _1 \phi _2 } \right\rangle \left| {\phi _5
\phi _6 } \right\rangle \left| {\phi _3 \phi _4 }
\right\rangle \left| {\phi _7 \phi _8 } \right\rangle$ &
$\left| {\phi _1 \phi _5 } \right\rangle \left| {\phi _2
\phi _6 } \right\rangle \left| {\phi _3 \phi _7 }
\right\rangle \left| {\phi _4 \phi _8 } \right\rangle  $ &
$\left| {\phi _1 \phi _6 } \right\rangle \left| {\phi _5
\phi _2 } \right\rangle \left| {\phi _3 \phi _8 }
\right\rangle \left| {\phi _7 \phi _4 } \right\rangle  $&
$\left| {\phi _1 \phi _6 } \right\rangle \left| {\phi _2
\phi _5 } \right\rangle \left| {\phi _3 \phi _8 }
\right\rangle \left| {\phi _4 \phi _7 } \right\rangle  $\\
\hline
 $s_1s_2s_3s_4$ & $\left| {1000} \right\rangle $        & $\left| {1001} \right\rangle $  & $\left| {1010} \right\rangle $ & $\left| {1011} \right\rangle $
\\ \hline
  $q_1q_2q_3q_4q_5q_6q_7q_8$ &$
\left| {\phi _1 \phi _2 } \right\rangle \left| {\phi _7
\phi _8 } \right\rangle \left| {\phi _5 \phi _6 }
\right\rangle \left| {\phi _3 \phi _4 } \right\rangle$ &
$\left| {\phi _1 \phi _7 } \right\rangle \left| {\phi _2
\phi _8 } \right\rangle \left| {\phi _5 \phi _3 }
\right\rangle \left| {\phi _6 \phi _4 } \right\rangle  $ &
$\left| {\phi _1 \phi _8 } \right\rangle \left| {\phi _7
\phi _2 } \right\rangle \left| {\phi _5 \phi _4 }
\right\rangle \left| {\phi _3 \phi _6 } \right\rangle  $&
$\left| {\phi _1 \phi _8 } \right\rangle \left| {\phi _2
\phi _7 } \right\rangle \left| {\phi _5 \phi _4 }
\right\rangle \left| {\phi _6 \phi _3 } \right\rangle  $\\
\hline
 $s_1s_2s_3s_4$ & $\left| {1100} \right\rangle $        & $\left| {1101} \right\rangle $  & $\left| {1110} \right\rangle $ & $\left| {1111} \right\rangle $
\\ \hline
$q_1q_2q_3q_4q_5q_6q_7q_8$ &$ \left| {\phi _1 \phi _2 }
\right\rangle \left| {\phi _7 \phi _8 } \right\rangle \left|
{\phi _3 \phi _4 } \right\rangle \left| {\phi _5 \phi _6
} \right\rangle$ & $\left| {\phi _1 \phi _7 } \right\rangle
\left| {\phi _2 \phi _8 } \right\rangle \left| {\phi _3
\phi _5 } \right\rangle \left| {\phi _4 \phi _6 }
\right\rangle  $ & $\left| {\phi _1 \phi _8 } \right\rangle
\left| {\phi _7 \phi _2 } \right\rangle \left| {\phi _3
\phi _6 } \right\rangle \left| {\phi _5 \phi _4 }
\right\rangle  $& $\left| {\phi _1 \phi _8 } \right\rangle
\left| {\phi _2 \phi _7 } \right\rangle \left| {\phi _3
\phi _6 } \right\rangle \left| {\phi _4 \phi _5 }
\right\rangle  $\\
 \hline
  \end{tabular}}
  \caption{The swap results of $s_1,s_2,s_3,s_4$ and $q_1,q_2,q_3,q_4,q_5,q_6,q_7,q_8$}
\end{table}

(3) Constructing a circuit ${\rm U}_m$ to estimate any two quantum states'
overlap for $m$ states $\left| {\phi _1 } \right\rangle ,\left|
{\phi _2 } \right\rangle ,...\left| {\phi _m } \right\rangle
$. If $2^{k - 1}  < m \le 2^k $, $2^k  - m$ $\left| 0 \right\rangle$
are added into $m$ states. In ${\rm U}_{n}(n=2^k)$, there are $2(k-1)$
ancillary qubits $s_1,s_2,...,s_{2(k-1)}$ and $2^k$ input states
$q_1,q_2,...,q_{2^k}$. The initial states are $s_i=\left| +
\right\rangle$ and $q_i=\left| {\phi _i }
\right\rangle(i=1,2,...,2^k)$. $q_1,q_2,...,q_{2^k}$ are divided
into four groups
$G_1=q_1,q_2,...,q_{2^{k-2}},G_2=q_{2^{k-2}+1},q_{2^{k-2}+2},...,q_{2(2^{k-2})},G_3=q_{2(2^{k-2})+1},q_{2(2^{k-2})+2},...,q_{3(2^{k-2})},G_4=q_{3(2^{k-2})+1},q_{3(2^{k-2})+2},...,q_{2^{k}}.$
There are also $(k - 1)2^{k - 1}$ CSWAP gates in ${\rm U}_n$. $s_1(s_2)$
is a control qubit on $\frac{{2^k }}{4}$ CSWAP gates to swap the
states in $G_2,G_3(G_2,G_4)$ according to $rule 1(rule 2)$. Then
the $2^{k-1}$ inputs of $G_1G_2(G_3G_4)$ can be used to construct a
circuit ${\rm U}_{2^{k-1}}$. After recursively constructing round by
round, $s_{(2(k - 1) - \frac{{2^k }}{4} + i)}, q_{4n - 3} ,q_{4n  -
2} ,q_{4n - 1} ,q_{4n } (i = 1,2,...,\frac{{2^k }}{4})$ can be used
as inputs of ${\rm U}_{4}$. $2^{k - 1}$ simple Swap Test for two quantum
states are places at the end of ${\rm U}_{2^{k}}$ to measure the overlap
for $(q_1,q_2)$, $(q_3,q_4)$,...,$(q_{2^k-1},q_{2^k})$.

\section{Simulation of multi-state Swap Test algorithm}
In this section, 8 random states $|\phi_i\rangle(i=1,2,...,8)$ are chosen as an example of a multi-state Swap Test algorithm and then the correctness of the algorithm on this example is verified by experiments on IBM quantum cloud platform.

For arbitrary two states $\left| {\phi _i } \right\rangle ,\left|
{\phi _j } \right\rangle (i,j = 1,2,...,8;i \ne j)$, the exact
overlap is denoted by $o(\left| {\phi _i } \right\rangle ,\left|
{\phi _j } \right\rangle )$, which can be calculated by the
definition of overlap $\left| {\left\langle {\phi,\psi }
\right\rangle } \right|^2$; And the estimated overlap is denoted by
$\widehat o(\left| {\phi _i } \right\rangle ,\left|{\phi _j }
\right\rangle )$, which can be obtained by executing simulated
experiments.

The simulated quantum circuit on IBM quantum cloud platform for 8-state Swap Test 
algorithm is given in Fig.(11), where $q_0q_1q_2q_3$ are the auxiliary control qubits that corresponds to
$s_1s_2s_3s_4$, $q_4q_5q_6q_7q_8q_9q_{10}q_{11}$ are the initial states that corresponds to $|\phi_1\rangle|\phi_2\rangle|\phi_3\rangle|\phi_4\rangle|\phi_5\rangle|\phi_6\rangle|\phi_7\rangle|\phi_8\rangle$ and $q_{12}q_{13}q_{14}q_{15}$ are the result qubits of simple
two-state Swap Test. The initial states $q_0q_1q_2q_3$ are $\left| +\right\rangle \left|  +  \right\rangle \left|  +  \right\rangle$.
Pair of normalized random numbers are generated by Python random
package and the initial states $q_4q_5q_6q_7q_8q_9q_{10}q_{11}$ in the experiment are $|\phi_1\rangle=[0.0864,
0.9963]$,$|\phi_2\rangle=[0.8391,
0.5440]$,$|\phi_3\rangle=[0.2202,
0.9755]$,$|\phi_4\rangle=[0.5486,
0.8361]$,$|\phi_5\rangle=[0.2607,
0.9654]$,$|\phi_6\rangle=[0.4164,
0.9091]$,$|\phi_7\rangle=[0.9981,
0.0609]$,$|\phi_8\rangle=[0.4658, 0.8849]$. And the measurement
results of this implementation through IBM Quantum Experience
(QISKIT) are given in Fig.(12),with 8192 runs. Meanwhile, the measurement times of $q_0q_1q_2q_3q_{12}q_{13}q_{14}q_{15}$ is shown in Appendix A.

\begin{figure}[htbp]
\includegraphics[height=7cm,width=12cm]{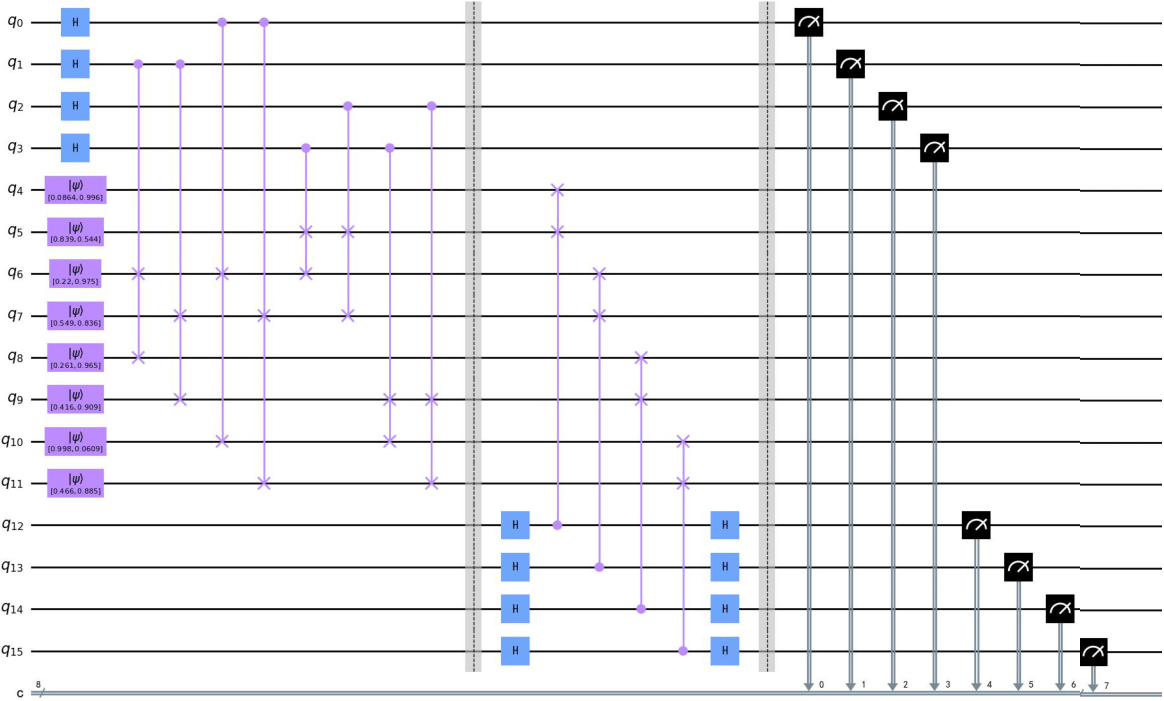}
\centering
\caption{The circuit for 8-state Swap Test algorithm}
\end{figure}

\begin{figure}[htbp]
    \includegraphics[height=10cm,width=18cm]{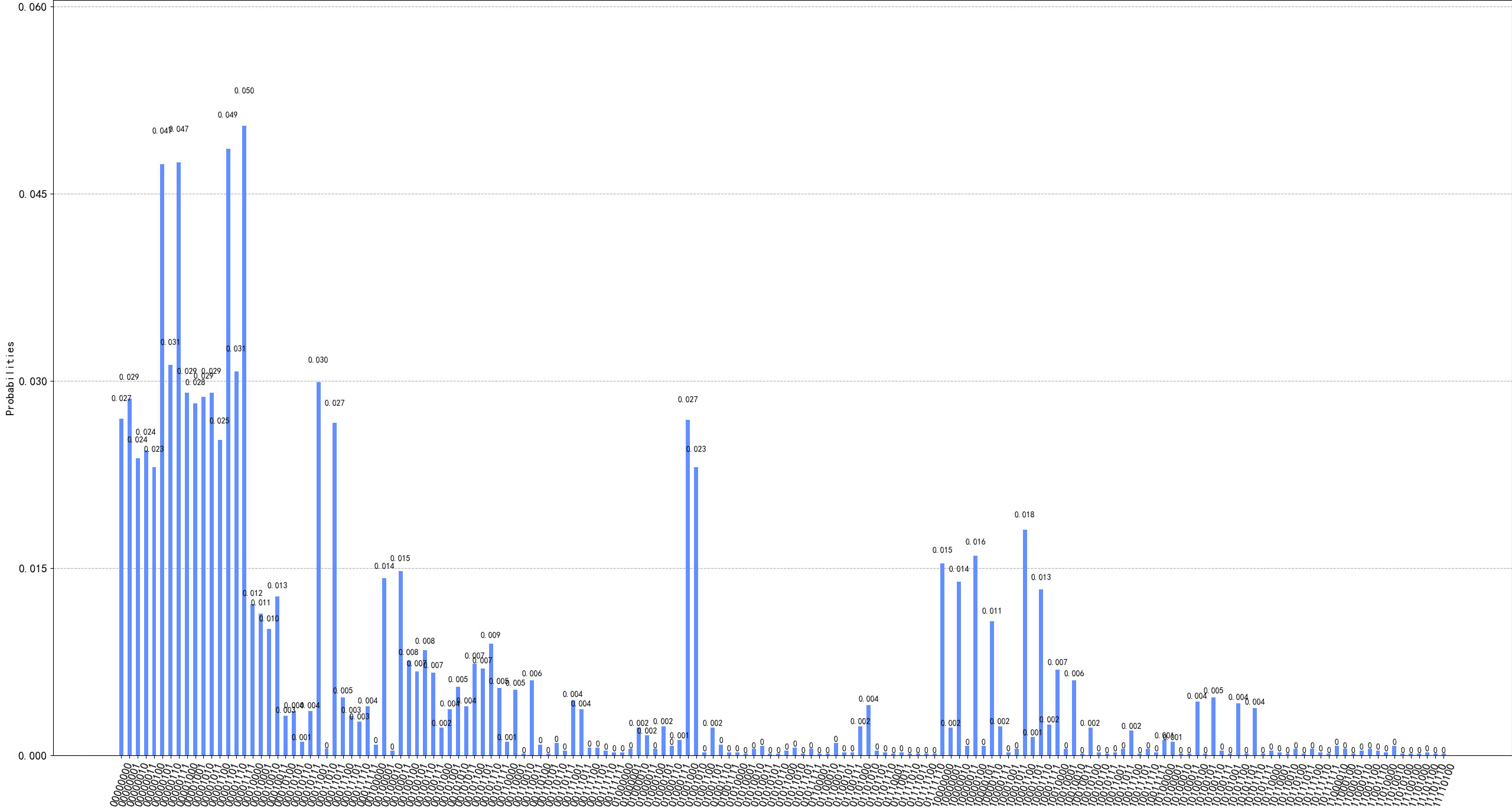}
      \centering
\caption{Result of implementation of 8-state Swap Test algorithm}
\end{figure}

The estimated overlap of $\left| {\phi _i } \right\rangle ,\left|
{\phi _j } \right\rangle (i,j = 1,2,...,8;i \ne j)$ can be
obtained according to the relation of $s_1,s_2,s_3,s_4$ and
$q_1,q_2,q_3,q_4,q_5,q_6,q_7,q_8$ in Table.2 and Appendix A. For
example, $s_1,s_2,s_3,s_4=\left| {0010} \right\rangle$ and
$s_1,s_2,s_3,s_4=\left| {0011} \right\rangle$, the measurement
result of $q_{15}$ is related to estimate values of $\left\langle
{\phi _6 } \right|\left. {\phi _7 } \right\rangle$. After
counting times of $q_{15}= \left| 1 \right\rangle$ $t_1=403$ and the
times of $q_{15}= \left| 0 \right\rangle$ $t_0=601$, $ \widehat
o(| {\phi _6 } \rangle,| {\phi _7 }
\rangle ) = {\frac{{2t_0 }}{{t_0  + t_1 }} - 1}  =
0.4441$. The exact overlap of arbitrary two states
$o(\left| {\phi _i } \right\rangle ,\left| {\phi _j }
\right\rangle )$ and the estimated overlap of arbitrary two
states $\widehat o(\left| {\phi _i } \right\rangle ,\left|
{\phi _j } \right\rangle )(i,j = 1,2,...,8;i \ne j)$ is shown in
Table.(4).
\begin{table}[htbp]\scriptsize
  \begin{tabular}{|l|l|l|l|l|l|l|l|l|l|l|}
  \hline
  $o(\left| {\phi _1 } \right\rangle
,\left| {\phi _2 } \right\rangle )$ & $\widehat o(\left| {\phi
_1 } \right\rangle ,\left| {\phi _2 } \right\rangle )$ &
$o(\left| {\phi _1 } \right\rangle ,\left| {\phi _3 }
\right\rangle )$ & $\widehat o(\left| {\phi _1 } \right\rangle
,\left| {\phi _3 } \right\rangle )$ & $o(\left| {\phi _1 }
\right\rangle ,\left| {\phi _4 } \right\rangle )$ &  $o(\left|
{\phi _1 } \right\rangle ,\left| {\phi _4 } \right\rangle )$ &
$o(\left| {\phi _1 } \right\rangle ,\left| {\phi _5 }
\right\rangle )$ & $\widehat o(\left| {\phi _1 } \right\rangle
,\left| {\phi _5 } \right\rangle )$
\\ \hline
0.3774&0.3959&0.9817&0.9880&0.7751&0.7590&0.9688&0.9610\\
\hline
 $o(\left| {\phi _1 } \right\rangle ,\left|
{\phi _6 } \right\rangle )$ & $\widehat o(\left| {\phi _1 }
\right\rangle ,\left| {\phi _6 } \right\rangle )$ & $o(\left|
{\phi _1 } \right\rangle ,\left| {\phi _7 } \right\rangle )$ &
$o(\left| {\phi _1 } \right\rangle ,\left| {\phi _7 }
\right\rangle )$ & $o(\left| {\phi _1 } \right\rangle ,\left|
{\phi _8 } \right\rangle )$ & $\widehat o(\left| {\phi _1 }
\right\rangle ,\left| {\phi _8 } \right\rangle )$ & $o(\left|
{\phi _2 } \right\rangle ,\left| {\phi _3 } \right\rangle )$ &
$\widehat o(\left| {\phi _2 } \right\rangle ,\left| {\phi _3 }
\right\rangle )$
\\ \hline
0.8868&0.8949&0.0215&0.0009&0.8497&0.8658&0.5118&0.5697\\
\hline
  $o(\left| {\phi _2 }
\right\rangle ,\left| {\phi _4 } \right\rangle )$ &  $o(\left|
{\phi _2 } \right\rangle ,\left| {\phi _4 } \right\rangle
)$&$o(\left| {\phi _2 } \right\rangle ,\left| {\phi _5 }
\right\rangle )$ & $\widehat o(\left| {\phi _2 } \right\rangle
,\left| {\phi _5 } \right\rangle )$ & $o(\left| {\phi _2 }
\right\rangle ,\left| {\phi _6 } \right\rangle )$ & $\widehat
o(\left| {\phi _2 } \right\rangle ,\left| {\phi _6 }
\right\rangle )$ & $o(\left| {\phi _2 } \right\rangle ,\left|
{\phi _7 } \right\rangle )$ &  $o(\left| {\phi _2 }
\right\rangle ,\left| {\phi _7 } \right\rangle )$
\\ \hline
0.8374&0.8476&0.5533&0.5393&0.7123&0.6913&0.7581&0.7714\\
\hline
  $o(\left| {\phi _2 } \right\rangle
,\left| {\phi _8 } \right\rangle )$ & $\widehat o(\left| {\phi
_2 } \right\rangle ,\left| {\phi _8 } \right\rangle )$ &
$o(\left| {\phi _3 } \right\rangle ,\left| {\phi _4 }
\right\rangle )$ & $\widehat o(\left| {\phi _3 } \right\rangle
,\left| {\phi _4 } \right\rangle )$ & $o(\left| {\phi _3 }
\right\rangle ,\left| {\phi _5 } \right\rangle )$ &  $o(\left|
{\phi _3 } \right\rangle ,\left| {\phi _5 } \right\rangle )$ &
$o(\left| {\phi _3 } \right\rangle ,\left| {\phi _6 }
\right\rangle )$ & $\widehat o(\left| {\phi _3 } \right\rangle
,\left| {\phi _6 } \right\rangle )$
\\ \hline
0.7607&0.7562&0.8768&0.8774&0.9982&0.9982&0.9574&0.9611\\
\hline

$o(\left| {\phi _3 } \right\rangle ,\left| {\phi _7 }
\right\rangle )$ & $\widehat o(\left| {\phi _3 } \right\rangle
,\left| {\phi _7 } \right\rangle )$ & $o(\left| {\phi _3 }
\right\rangle ,\left| {\phi _8 } \right\rangle )$ & $o(\left|
{\phi _3 } \right\rangle ,\left| {\phi _8 } \right\rangle )$ &
$o(\left| {\phi _4 } \right\rangle ,\left| {\phi _5 }
\right\rangle )$ & $\widehat o(\left| {\phi _4 } \right\rangle
,\left| {\phi _5 } \right\rangle )$ & $o(\left| {\phi _4 }
\right\rangle ,\left| {\phi _6 } \right\rangle )$ & $\widehat
o(\left| {\phi _4 } \right\rangle ,\left| {\phi _6 }
\right\rangle )$
\\ \hline
0.0779&0.1093&0.9325&0.9516&0.9028&0.9104&0.9773&0.9739\\
\hline

 $o(\left| {\phi _4 } \right\rangle ,\left| {\phi _7 }
\right\rangle )$ &  $o(\left| {\phi _4 } \right\rangle ,\left|
{\phi _7 } \right\rangle )$ & $o(\left| {\phi _4 }
\right\rangle ,\left| {\phi _8 } \right\rangle )$ & $\widehat
o(\left| {\phi _4 } \right\rangle ,\left| {\phi _8 }
\right\rangle )$ & $o(\left| {\phi _5 } \right\rangle ,\left|
{\phi _6 } \right\rangle )$ & $\widehat o(\left| {\phi _5 }
\right\rangle ,\left| {\phi _6 } \right\rangle )$ & $o(\left|
{\phi _5 } \right\rangle ,\left| {\phi _7 } \right\rangle )$ &
$o(\left| {\phi _5 } \right\rangle ,\left| {\phi _7 }
\right\rangle )$
\\ \hline
0.3582&0.4323&0.9908&1.0&0.9727&0.9727&0.1017&0.0260\\
\hline $o(\left| {\phi _5 } \right\rangle ,\left| {\phi _8 }
\right\rangle )$ & $\widehat o(\left| {\phi _5 } \right\rangle
,\left| {\phi _8 } \right\rangle )$ & $o(\left| {\phi _6 }
\right\rangle ,\left| {\phi _7 } \right\rangle )$ & $\widehat
o(\left| {\phi _6 } \right\rangle ,\left| {\phi _7 }
\right\rangle )$ & $o(\left| {\phi _6 } \right\rangle ,\left|
{\phi _8 } \right\rangle )$ & $o(\left| {\phi _6 }
\right\rangle ,\left| {\phi _8 } \right\rangle )$ &   $o(\left|
{\phi _7 } \right\rangle ,\left| {\phi _8 } \right\rangle )$ &
$\widehat o(\left| {\phi _7 } \right\rangle ,\left| {\phi _8 }
\right\rangle )$
\\ \hline
0.9519&0.9602&0.2218&0.1972&0.9970&0.9960&0.2691&0.2927\\
\hline
  \end{tabular}
  \caption {The exact overlap value of arbitrary two states
$o(\left| {\phi _i } \right\rangle ,\left| {\phi _j }
\right\rangle )$ and the estimate overlap value of arbitrary two
states $\widehat o(\left| {\phi _i } \right\rangle ,\left|
{\phi _j } \right\rangle )(i,j = 1,2,...,8;i \ne j)$}
\end{table}
The scatter diagram is shown in Fig.(13), where the estimate overlap
value $\widehat o(\left| {\phi _i } \right\rangle ,\left|
{\phi _j } \right\rangle )$ is X-axis and the exact overlap
value $o(\left| {\phi _i } \right\rangle ,\left| {\phi _j }
\right\rangle )$ is Y-axis. It can be seen from Fig.(13) that most
of the points fall near the diagonals of the picture, i.e. the line
$y=x$, indicating that the estimate overlap results are close to the
exact overlap results.

\begin{figure}[htbp]
    \includegraphics[height=5cm,width=5cm]{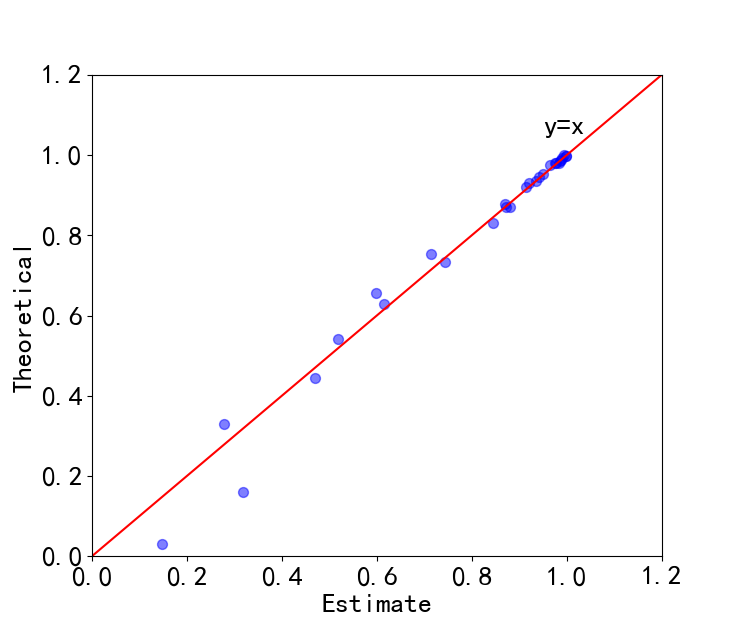}
      \centering
\caption{The relation of $o(\left| {\phi _i } \right\rangle
,\left| {\phi _j } \right\rangle )$ and $\widehat o(\left|
{\phi _i } \right\rangle ,\left| {\phi _j } \right\rangle )$}
  \label{fig:1}       
\end{figure}

9 data sets of the initial states $D_1,D_2,D_3,D_4,D_5,D_6,D_7,D_8,D_9$ are generated by Python random package to be simulated in this experiment and their initial values are shown in Appendix B. And the scatter diagrams of estimated overlaps and exact overlaps of these 9 sets are shown in Fig.(14), where the estimated overlaps is X-axis and the exact overlaps is Y-axis. It can be seen from Fig.6 that most of the points fall near the diagonals of the picture, i.e., the line $y=x$, indicating that the estimated overlaps are close to the exact overlaps.

\begin{figure}[htbp]
    \centering
    \subfigure[$D_1$]{
        \includegraphics[width=1.8in]{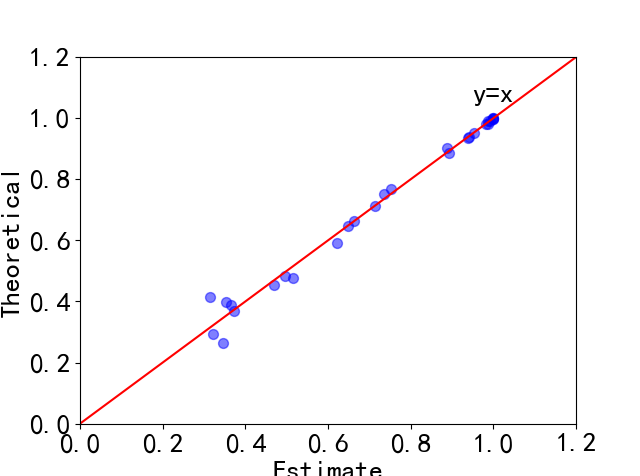}
    }
 \subfigure[$D_2$]{
    \includegraphics[width=1.8in]{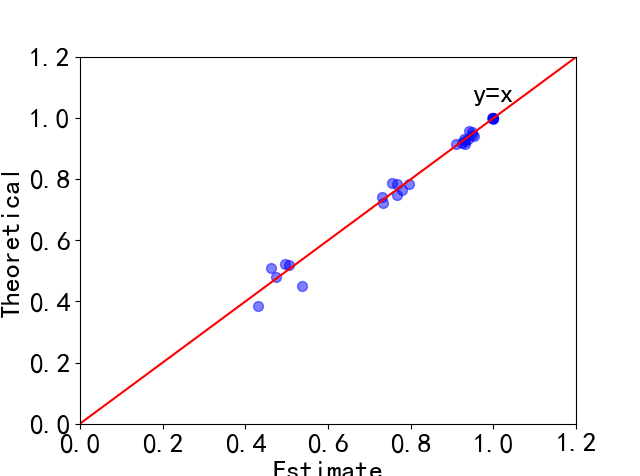}
    }
     \subfigure[$D_3$]{
    \includegraphics[width=1.8in]{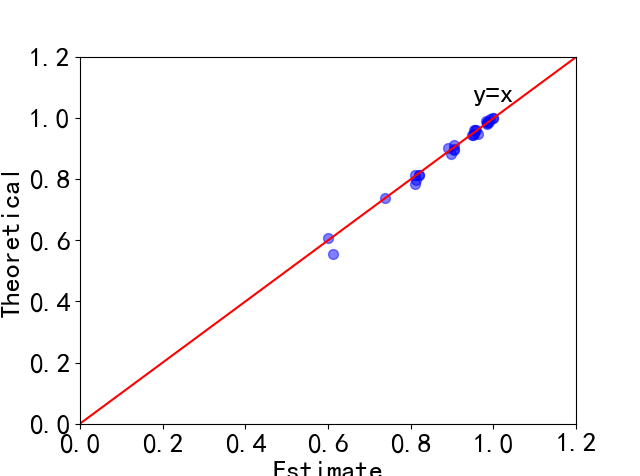}
    }
    \quad
    \subfigure[$D_4$]{
        \includegraphics[width=1.8in]{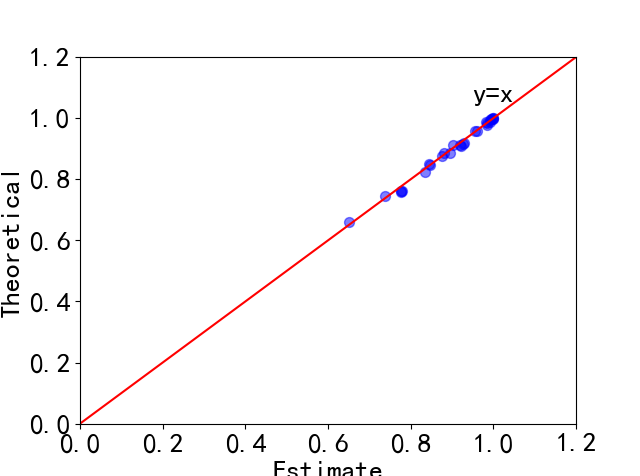}
    }
    \subfigure[$D_5$]{
    \includegraphics[width=1.8in]{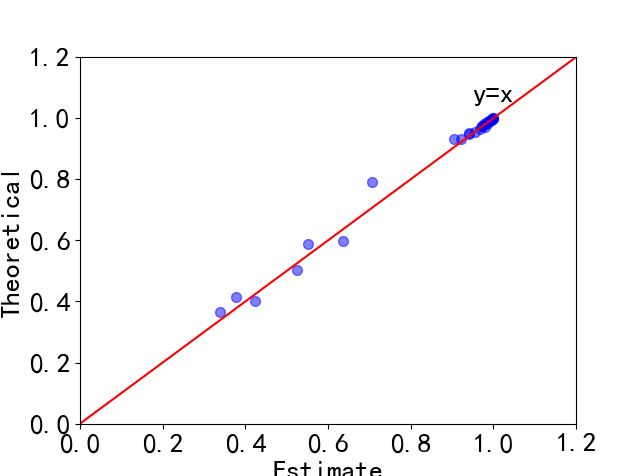}
    }
     \subfigure[$D_6$]{
    \includegraphics[width=1.8in]{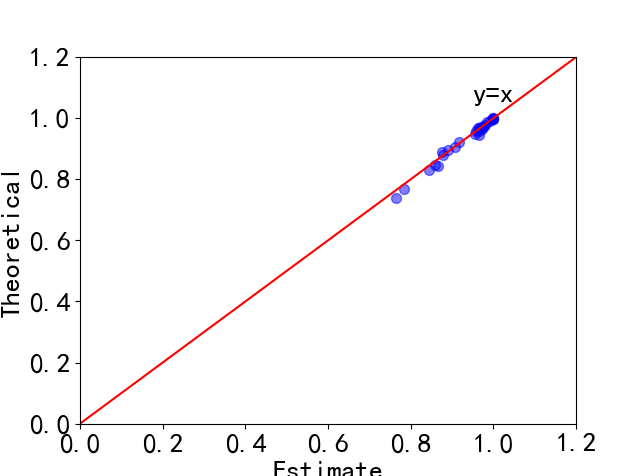}
    }    \quad
    \subfigure[$D_7$]{
        \includegraphics[width=1.8in]{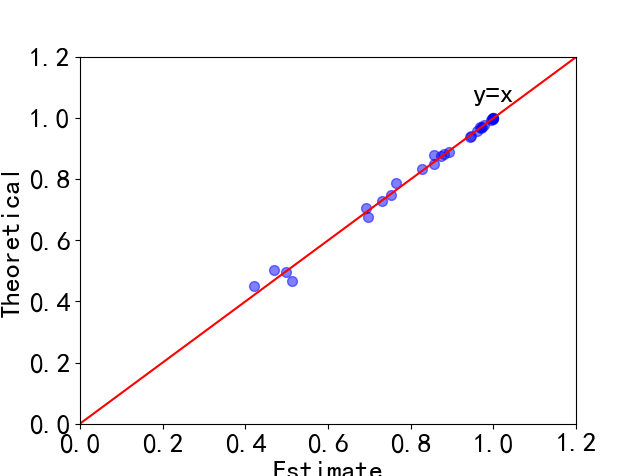}
    }
    \subfigure[$D_8$]{
    \includegraphics[width=1.8in]{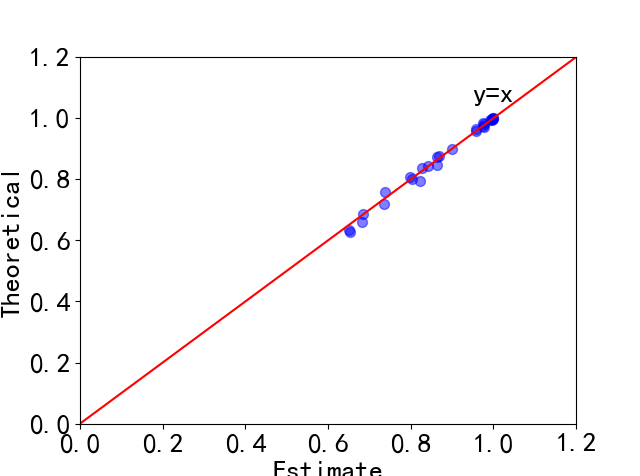}
    }
     \subfigure[$D_9$]{
    \includegraphics[width=1.8in]{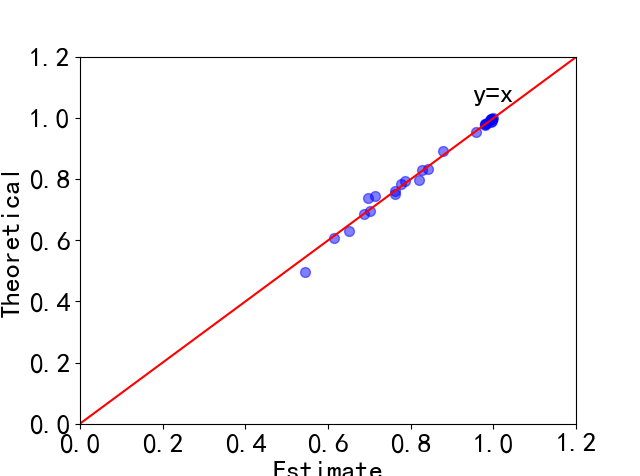}
        }
    \caption{The scatter diagrams of 9 groups quantum states}
    \label{fig.1}
\end{figure}
\section{Analysis}

In this section, we analyze the correctness and performance of our multi-state Swap Test algorithm in theory.

\subsection{Correctness analysis}
There is a transformation ${\rm U}_n$, which can take arbitrary unknown
states $\left| {\phi _1 } \right\rangle ,\left| {\phi _2 }
\right\rangle ,...\left| {\phi _n } \right\rangle (n = 2^k (k >
1))$ as input states and $2(k-1)$ $\left|  +  \right\rangle$ as
ancillaries. In ${\rm U}_n$, each pair $\left| {\phi _{2i-1} }
\right\rangle ,\left| {\phi _{2i} } \right\rangle (i =
1,2,...,\frac{n}{2})$ can be labeled by ancillaries $\left| {A_{(2i
- 1)2i} } \right\rangle$ by Lemma 4.1.

\textbf{Lemma 4.1} Let $n = 2^k (k > 1)$, there exists a
unitary ${\rm U}_n$ that maps

\begin{equation}
  U_n:|+\rangle^{d_n}\otimes_{i=1}^n|\phi_i\rangle\rightarrow|y_{ij}\rangle+|1_{ij}\rangle|G_{ij}^{1'}\phi_i\phi_jG_{ij}^{1}\rangle+...+|m_{ij}\rangle|G_{ij}^{m'}\phi_i\phi_jG_{ij}^m\rangle,
\end{equation}

where $|\phi_i\rangle,|\phi_j\rangle$ are arbitrary two input quantum states, 
and they are 
in adjacent quantum registers, i.e. in the quantum registers $(q_{2i-1},q_{2i}),i\in\{1,2,...,{n\over2}\}$.
$|G_{ij}'\rangle,|G_{ij}\rangle,|y_{ij}\rangle$ are garbage states which is irrelevant to our 
target states, i.e. $|\phi_i\rangle,|\phi_j\rangle$.
$|1_{ij}\rangle,...,|m_{ij}\rangle$ represents the state in the first $d_n$ quantum registers.
That means we can get all pairs $|\phi_i\rangle|\phi_j\rangle,i\neq j$ by measuring the specific
quantum registers $(q_{2i-1},q_{2i}),i\in\{1,2,...,{n\over2}\}$ in the quantum circuit.

\textbf{Constructive proof of Lemma 4.1} According to scientific induction, we assume 
that Lemma 4.1 is correct at first. Let the $n$ quantum registers divided into 4 groups 
$Q_1^1=(q_1,...,q_{n/4})$,$Q_1^2=(q_{n/4+1},...,q_{n/2})$,
$Q_1^3=(q_{n/2+1},...,q_{3n/4})$,$Q_1^4=(q_{3n/4+1},...,q_{n})$. Furthermore, we get 4 quantum states
$|\Phi_1^1\rangle=|\phi_1...\phi_{n/4}\rangle$,$|\Phi_1^2\rangle=|\phi_{n/4+1},...,\phi_{n/2}\rangle$,
$|\Phi_1^3\rangle=|\phi_{n/2+1},...,\phi_{3n/4}\rangle$,$|\Phi_1^4\rangle=|\phi_{3n/4+1},...,\phi_{n}\rangle$.
Applying ${\rm U}_{n/2}$ to $(Q_1^1,Q_1^2)$ and $(Q_1^3,Q_1^4)$ respectively, we can get all pairs of input states
$|\Phi_1^1\Phi_1^2\rangle$, i.e. $|\phi_1\rangle,...,|\phi_{n/2}\rangle$, and the same is true for the other 
${\rm U}_{n/2}$. If we apply a unitary ${\rm U}_4$ on $(Q_1^1,Q_1^2,Q_1^3,Q_1^4)$ before the ${\rm U}_{n/2}$, 
we will get all pairs of input states $|\phi_1\rangle,...,|\phi_{n}\rangle$. For example, we want to 
get a pair $|\phi_a\phi_b\rangle,a\in[1,{n\over4}],b\in[{n\over2}+1,{3n\over4}]$. $|\Phi_1^1\Phi_1^3\rangle$
will be stored in $(Q_1^1,Q_1^2)$ if the ancillary qubits of ${\rm U}_4$ is $|01\rangle$, and the pair 
$|\phi_a\phi_b\rangle$ might be measured after ${\rm U}_{n/2}$.

Similarly, the correctness of ${\rm U}_{n/2}$ can be provide by using ${\rm U}_{n/4}$ and ${\rm U}_4$, and so on.
Finally, the correctness of ${\rm U}_{4}$ is shown in Table.1. So Lemma 4.1 is proved.
For the case that $n$ cannot be written in the form of $2^k$, we can add $2^k-n$ copy of 
any input state as the rest input.

So far, we have obtained all pairs of $n$ input quantum states, and can get
each pair exactly according to the measurement results of auxiliary qubits.
Therefore, we only need to apply the Swap Test in all
specific registers $(q_1,q_2),...,(q_{n-1},q_n)$ to get the overlap between two input quantum
states. As shown in Figure(16), the measurement results of $2(k-1)$ auxiliaries determine the quantum
state in each quantum register uniquely. Because it has been proved that all pairs of input quantum states
have the opportunity to be measured in specific registers, all overlap estimates can be obtained after enough measurements.

\subsection{Performance analysis}
In this section, we compare our scheme and the scheme Swap Test for an arbitrary number of quantum states given by Gitiaux(referred to as SAN scheme later in this article).

\subsubsection{Precision}

The precision of estimation is determined by the occurrences of each overlap with the total number of runs given, thus we can use this value ,denoted by m to represent the precision.
For the SAN scheme, only one measurement can be obtained each time the quantum circuit
is run. Therefore, in the case of $n=2^k$ quantum state inputs, if the count of
quantum circuit is run for $N$ times, each overlap is estimated to occur $2N\over {n (n-1)} $
times in average, i.e., $m_1={2N\over {n (n-1)}}$. For our scheme, one execution
contains $n\over 2 $ Swap Test operations. So in the same case, we have the precision $m_2={N\over {n-1}}$,which means higher precision.

  \begin{figure}[htbp]
    \includegraphics[height=5cm,width=6cm]{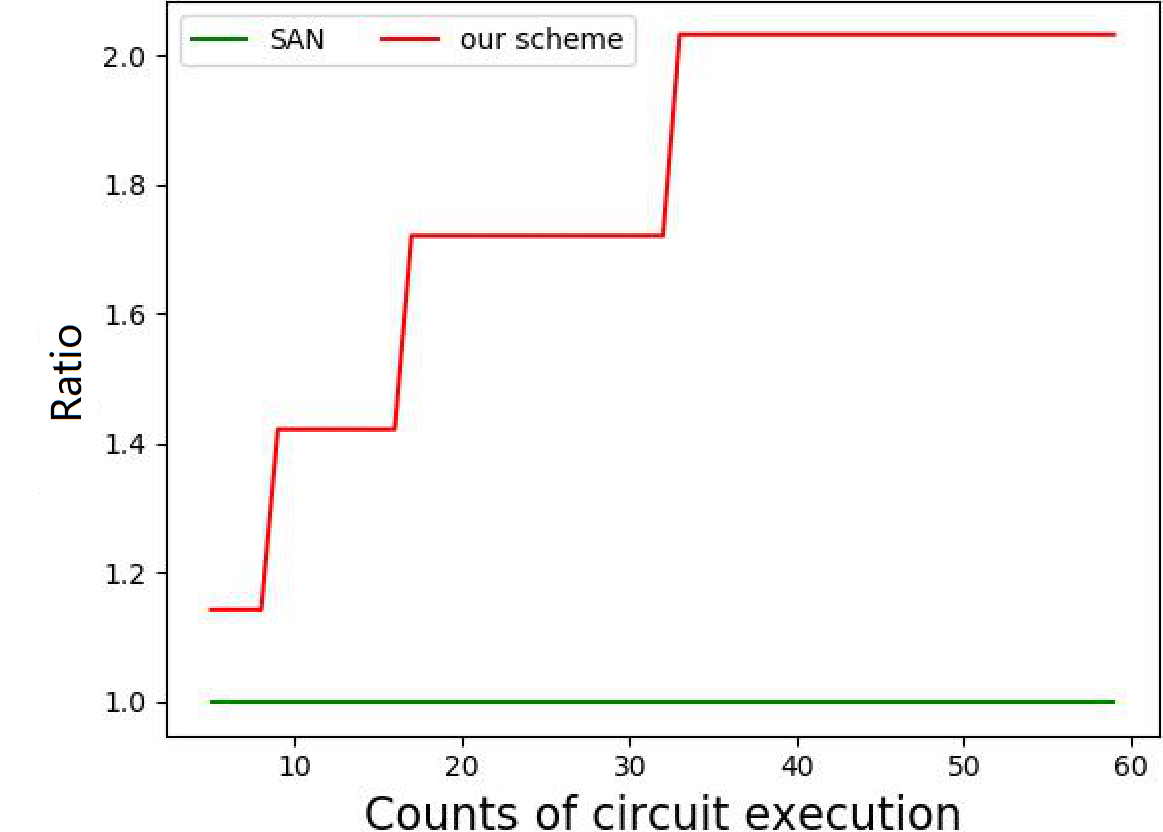}
    \centering
  \caption{The precision for different schemes(expressed in terms of ratio)}
  \end{figure}

Fig.(15) shows the ratio of the average precision of our scheme with respect to SAN. Taking the number of the SAN scheme $m_1$ as the standard, it can be seen that with the increase of the circuits depth, the ratio between the new scheme and SAN scheme $m_2\over m_1$ increases gradually, indicating that our precision advantage becomes gradually more pronounced as the circuit depth increases.The stepwise growth is due to the fact that we must apply such $k$ that $2^{k-1} <n \leq 2^k$. It should be noted that if the probability of occurrence of each quantum state is taken into account, the estimation of different overlaps will have different precision.This is because the number of recurrences of different pairs varies according to the table, which means that some pairs may occur more frequently. However, this only depends on the artificial choices we make when coding, and the total effect does not have a large impact as the number increases.

\subsubsection{Space complexity}

The space complexity of ${\rm U}_n$ consists of the count of CSWAP gates
and the number of ancillaries, where $n=2^k$ and $k=log_2n$. First
of all, we notice that the count of CSWAP gates $c_n$ follows the
recursive relation:
\begin{equation}
c_n=2c_{n/2}+{n\over2}
\end{equation}
where the first term comes from two ${\rm U}_{n/2}$ and the second term
comes from a ${\rm U}_4$. The general term formula $c_n=nk$ can be
obtained by simplifying the sequence relation. If considering the
CSWAP gates come from the following Swap Test operation, we
have$c_n=n(k+{1\over2})$. Similarly, for SAN scheme, we can get the
general term formula for the number of CSWAP gates $c_n'=3(n-1)$,
which is shown in the right panel of Fig.(16). Furthermore, if we
concentrate on the count of CSWAP gates, the complexity of ${\rm U}_{n}$
in our scheme is$O(n\log n)$ and in SAN scheme is ${\rm
O}(n)$. Obviously, our scheme uses more CSWAP gates and has higher
complexity than the SAN scheme.

As for the number of ancillaries, we have the relation:
  \begin{equation}
    d_n=d_{n-1}+2
  \end{equation}
which means each time the number of input quantum states is doubled,
2 additional auxiliaries are required. It can be rewritten as
$d_n=2k-2$, and the general term formula for SAN scheme the number
of ancillaries is $d_n'=3k-3$. Obviously, our scheme uses fewer
ancillaries than the SAN scheme, and both complexities are ${\rm
O}(\log n)$. The relationship between the number of auxiliaries and
the number of input quantum states (assuming that all input quantum
states are single quantum states) is shown in Figure(18). As shown
in the left panel. the red line and the green line represent our
scheme and SAN scheme respectively. It can be seen that the number
of auxiliaries required by our new scheme is less than that of the
SAN scheme.

 \begin{figure}[htbp]
\includegraphics[height=5cm,width=12cm]{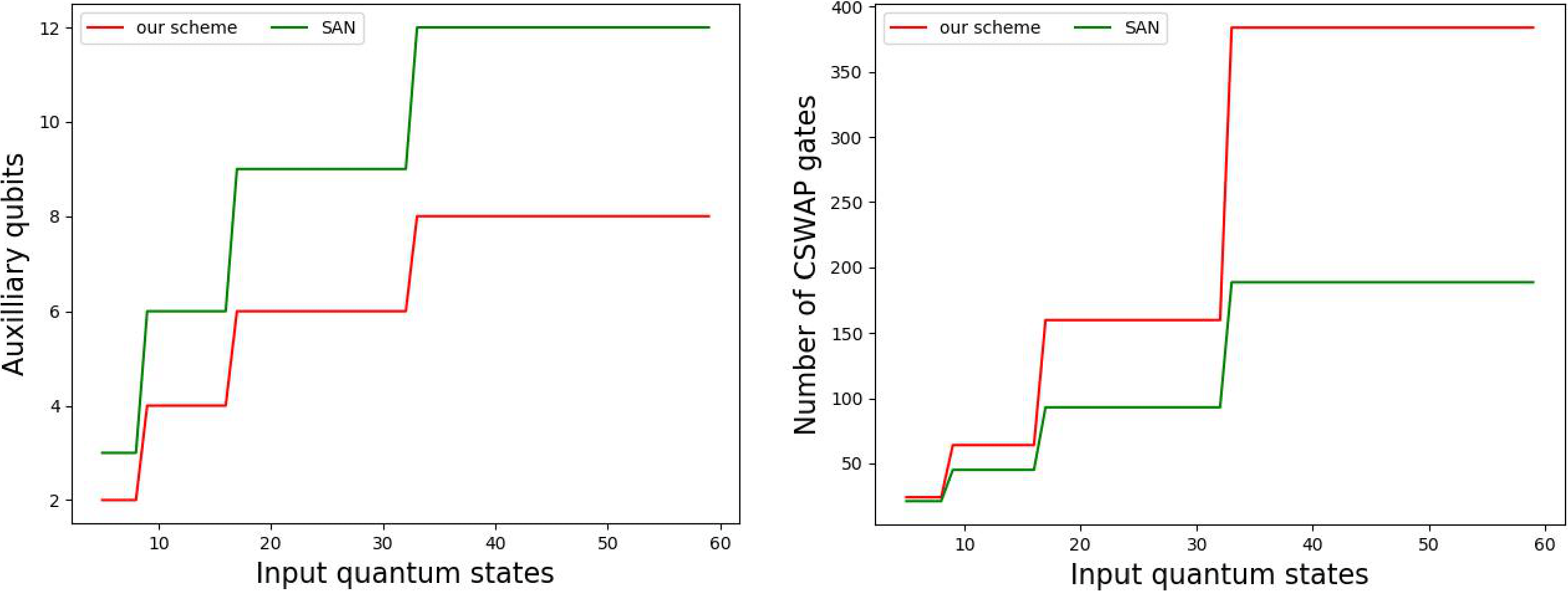}
\centering
\caption{The complexity of two schemes. The left panel shows the count of ancillaries
of two schemes, and the complexities of CSWAP gates is shown in the right panel.}
\end{figure}

If we consider the additional auxiliaries brought by the subsequent Swap Test of the two Swap Test schemes, we can reduce it by using the Swap Test circuit optimization scheme mentioned in section 2 to reduce the auxiliaries through classical calculation. At this time, the total number of qubits required by our scheme will be less than that required by the SAN scheme, which means less resource consumption.

To sum up, our scheme provides a better estimation under the same
count of circuit runs. At the same time, the number of auxiliaries
required in our scheme is less than that of the SAN scheme. Although
the complexity of the count of CSWAP gates in the SAN scheme is ${\rm
O}(n)$ which is less than that in our scheme $O(n\log n)$,
When the circuits run, they are processed in parallel and do not
affect the total time, so we trade a certain space cost for the
desired time cost.

\section{Summary}
In this paper, we propose a new scheme for estimating overlap
between all pair of
  input quantum states. Given arbitrary number of input quantum states $|\phi_1\rangle,...|\phi_n\rangle$.
  the scheme provide a way to calculate the overlap $|\langle\phi_i|\phi_j\rangle|^2$ for
  all pair of input quantum states where $i<j\ i,j\in{1,2,...,n}$.

  Our scheme is based on the scheme given by \cite{19}. The basic principle is we concentrate a unitary
  ${\rm U}_n$ according to $n$ input quantum states $|\phi_1\rangle,...,|\phi_n\rangle$, which is used to obtain
  a specific superposition state, so that we can get all pairs of input quantum states at a specific position
  $(q_{2n-1},q_{2n}),n\in{1,2,...,{n\over2}}$. In order to decode the measurement results, we use $O(\log n)$
  auxiliary qubits, their measurement results reveal what the quantum state in the register is. Finally,
  the results are obtained by performing Swap Test on these positions to estimate the value of overlap between
  each pair. A total of $O(n\log n)$ CSWAP gates are used in the quantum circuit. We carry out
  experiments on IBM cloud platform, and give a exact analysis to
  verify the correctness of our scheme. At the same time, we analyze the complexity and accuracy of
  our scheme. Compared with the original scheme \cite{19}, our scheme uses fewer auxiliary qubits
  and has higher accuracy, but the line depth (the number of CSWAP gates) will be higher than the
  original scheme. Therefore, if it is possible to find a way to reduce the use of cswap gates, the
  performance of the scheme will be further improved.

\ack
This work was supported in part by the 2019 National Social
Science Foundation Art Major Project, Network Culture Security
Research, under Grant 19zd12, in part by the High-Quality and
Cutting-Edge Disciplines Construction Project for Universities in
Beijing (Internet Information, Communication University of China),
and in part by the Fundamental Research Funds for the Central
Universities, and by Equipment Pre-research Field Fund under Grant
61403110320.

\section*{Appendix}
\subsection*{A }
\begin{table}[H]\scriptsize
\centering
    \begin{tabular}{|l|l|l|l|l|l|}
    \hline
    $|q_0q_1q_2q_3q_{12}q_{13}q_{14}q_{15}\rangle$ & counts & $|q_0q_1q_2q_3q_{12}q_{13}q_{14}q_{15}\rangle$ & counts & $|q_0q_1q_2q_3q_{12}q_{13}q_{14}q_{15}\rangle$ & counts \\ \hline
    $|10110000\rangle$ & 398 & $|01100101\rangle$ & 38  & $|00111000\rangle$ & 38  \\ \hline
    $|01010000\rangle$ & 235 & $|00110000\rangle$ & 207 & $|00000100\rangle$ & 7   \\ \hline
    $|10011000\rangle$ & 245 & $|11011100\rangle$ & 30  & $|01000110\rangle$ & 1   \\ \hline
    $|00110100\rangle$ & 60  & $|11101100\rangle$ & 3   & $|00001011\rangle$ & 2   \\ \hline
    $|01001000\rangle$ & 83  & $|10100100\rangle$ & 55  & $|10000010\rangle$ & 4   \\ \hline
    $|00000000\rangle$ & 221 & $|01110101\rangle$ & 31  & $|01111000\rangle$ & 22  \\ \hline
    $|10011100\rangle$ & 36  & $|00110001\rangle$ & 148 & $|01101000\rangle$ & 9   \\ \hline
    $|11100010\rangle$ & 10  & $|00110101\rangle$ & 34  & $|00100011\rangle$ & 4   \\ \hline
    $|11100000\rangle$ & 389 & $|10000101\rangle$ & 9   & $|10101000\rangle$ & 29  \\ \hline
    $|01100000\rangle$ & 256 & $|11010100\rangle$ & 32  & $|01001011\rangle$ & 6   \\ \hline
    $|10100000\rangle$ & 388 & $|00101101\rangle$ & 4   & $|00111100\rangle$ & 5   \\ \hline
    $|10000100\rangle$ & 116 & $|11100001\rangle$ & 19  & $|10111010\rangle$ & 1   \\ \hline
    $|01010100\rangle$ & 45  & $|01110011\rangle$ & 3   & $|00010110\rangle$ & 19  \\ \hline
    $|01000001\rangle$ & 114 & $|10001001\rangle$ & 4   & $|10000001\rangle$ & 18  \\ \hline
    $|11010000\rangle$ & 238 & $|01000010\rangle$ & 18  & $|01000011\rangle$ & 6   \\ \hline
    $|11101000\rangle$ & 29  & $|00100100\rangle$ & 62  & $|00110111\rangle$ & 2   \\ \hline
    $|00010010\rangle$ & 220 & $|00100101\rangle$ & 35  & $|01111001\rangle$ & 4   \\ \hline
    $|01110100\rangle$ & 73  & $|00111001\rangle$ & 16  & $|01000100\rangle$ & 3   \\ \hline
    $|10001100\rangle$ & 43  & $|11110001\rangle$ & 20  & $|00000101\rangle$ & 10  \\ \hline
    $|00101000\rangle$ & 26  & $|10110100\rangle$ & 57  & $|11110101\rangle$ & 1   \\ \hline
    $|11111010\rangle$ & 44  & $|11001100\rangle$ & 49  & $|00001100\rangle$ & 9   \\ \hline
    $|01010010\rangle$ & 189 & $|00100010\rangle$ & 4   & $|11010001\rangle$ & 4   \\ \hline
    $|11111010\rangle$ &  4  & $|01100001\rangle$ & 88  & $|11000110\rangle$ & 8   \\ \hline
    $|11111000\rangle$ & 32  & $|10100010\rangle$ & 19  & $|10011001\rangle$ & 2   \\ \hline
    $|11000000\rangle$ & 200 & $|01010110\rangle$ & 33  & $|01111110\rangle$ & 1   \\ \hline
    $|10101010\rangle$ &  1  & $|00111101\rangle$ & 4   & $|10100001\rangle$ & 6   \\ \hline
    $|00010000\rangle$ & 238 & $|10110010\rangle$ & 18  & $|11000001\rangle$ & 6   \\ \hline
    $|00100001\rangle$ & 131 & $|10010100\rangle$ & 30  & $|01011000\rangle$ & 4   \\ \hline
    $|11011000\rangle$ & 218 & $|01011010\rangle$ & 5   & $|00011010\rangle$ & 3   \\ \hline
    $|10001000\rangle$ & 93  & $|00101001\rangle$ & 18  & $|10000110\rangle$ & 1   \\ \hline
    $|00100000\rangle$ & 189 & $|01101001\rangle$ & 2   & $|10111001\rangle$ & 1   \\ \hline
    $|11100100\rangle$ & 54  & $|01100100\rangle$ & 69  & $|01000101\rangle$ & 1   \\ \hline
    $|00000001\rangle$ & 126 & $|01110110\rangle$ & 2   & $|00001101\rangle$ & 3   \\ \hline
    $|11000100\rangle$ & 121 & $|00000010\rangle$ & 2   & $|11001010\rangle$ & 6   \\ \hline
    $|01000000\rangle$ & 195 & $|10110001\rangle$ & 12  & $|01011110\rangle$ & 1   \\ \hline
    $|00001000\rangle$ & 99  & $|11100110\rangle$ & 2   & $|11110010\rangle$ & 2   \\ \hline
    $|11110000\rangle$ & 413 & $|00010100\rangle$ & 18  & $|11011001\rangle$ & 4   \\ \hline
    $|11001000\rangle$ & 104 & $|10111000\rangle$ & 26  & $|11100101\rangle$ & 3   \\ \hline
    $|01110001\rangle$ & 109 & $|01100010\rangle$ & 6   & $|00101111\rangle$ & 1   \\ \hline
    $|01001001\rangle$ & 49  & $|01110010\rangle$ & 7   & $|11111100\rangle$ & 2   \\ \hline
    $|10010000\rangle$ & 231 & $|00101100\rangle$ & 7   & $|00101011\rangle$ & 1   \\ \hline
    $|10000000\rangle$ & 234 & $|10111100\rangle$ & 5   & $|01001100\rangle$ & 1   \\ \hline
    $|01110000\rangle$ & 252 & $|01101100\rangle$ & 8   & $|01001101\rangle$ & 1   \\ \hline
    $|00001001\rangle$ & 56  & $|01100011\rangle$ & 3   & $|00110011\rangle$ & 4   \\ \hline
    $|11000010\rangle$ & 13  & $|01111101\rangle$ & 2   & $|10001110\rangle$ & 1   \\ \hline
    $|10110110\rangle$ & 3   & $|10001101\rangle$ & 2   & $|11000101\rangle$ & 1   \\ \hline
    $|11001001\rangle$ & 1   & $|01001010\rangle$ & 4   & $|10001010\rangle$ & 1   \\ \hline
    $|00010111\rangle$ & 1   & $|01101010\rangle$ & 1   & $|00001010\rangle$ & 2   \\ \hline
    $|11001110\rangle$ & 2   & $|00111110\rangle$ & 1   & $|00110010\rangle$ & 2   \\ \hline
    $|10100110\rangle$ & 2   & $|10101100\rangle$ & 1   & $|11101001\rangle$ & 1   \\ \hline
    $|00100111\rangle$ & 1   & $|01111100\rangle$ & 3   & $|11011110\rangle$ & 1   \\ \hline
    $|01110111\rangle$ & 1   & $|11111101\rangle$ & 1   & $|10010001\rangle$ & 2   \\ \hline
    $|10110101\rangle$ & 1   & $|11111001\rangle$ & 2   & $|10010101\rangle$ & 1   \\ \hline
    $|10100101\rangle$ & 1   & $|10100011\rangle$ & 1   & $|11011101\rangle$ & 1   \\ \hline
    \end{tabular}
     \caption{The counts of different measurement result of ancillary qubits $q_0 ,q_1 ,q_2 ,q_3,q_{12}
,q_{13},q_{14},q_{15}$}
\end{table}

\subsection*{B }

\begin{table}[H]
\centering
    \begin{tabular}{|l|l|l|l|l|l|}
    \hline

$D_1:$&$D_2:$&$D_3:$\\ \hline

$\begin{array}{l}
 |\phi_1\rangle=[0.1340, 0.9910]\\|\phi_2\rangle=[0.5666,
0.8240] \\
 |\phi_3\rangle=[0.1262, 0.9920]\\|\phi_4\rangle=[0.2899,
0.9571] \\
 |\phi_5\rangle=[0.9746, 0.2238] \\|\phi_6\rangle=[0.4602,
0.8878] \\
|\phi_7\rangle=[0.9813, 0.1926]\\|\phi_8\rangle=[0.9695,
0.2452] \\
 \end{array}
 $
 &
$\begin{array}{l}
 |\phi_1\rangle=[0.9697, 0.2443]\\|\phi_2\rangle=[0.8136,
0.5814] \\
 |\phi_3\rangle=[0.2681, 0.9634]\\|\phi_4\rangle=[0.1963,
0.9806] \\
 |\phi_5\rangle=[0.5862, 0.8102]\\|\phi_6\rangle=[0.2324,
0.9726] \\
|\phi_7\rangle=[0.9565, 0.2917]\\|\phi_8\rangle=[0.5461,
0.8377]\\
 \end{array}
 $
 &
$\begin{array}{l}
|\phi_1\rangle=[0.8240,0.5665]\\|\phi_2\rangle=[0.0398, 0.9992]\\

|\phi_3\rangle=[0.3343, 0.9425]\\|\phi_4\rangle=[0.4898,
0.8719]\\

|\phi_5\rangle=[0.8140, 0.5808]\\|\phi_6\rangle=[0.7047,
0.7095]\\

|\phi_7\rangle=[0.3366, 0.9416]\\|\phi_8\rangle=[0.6131,
0.7900]\\
 \end{array}
 $
 \\ \hline

$D_4:$&$D_5:$&$D_6:$ \\ \hline

$\begin{array}{l}

|\phi_1\rangle=[0.2945, 0.9557]\\|\phi_2\rangle=[0.6896,
0.7242]\\

|\phi_3\rangle=[0.1983, 0.9801]\\|\phi_4\rangle=[0.3120,
0.9501]\\

|\phi_5\rangle=[0.6402, 0.7682]\\|\phi_6\rangle=[0.0190,
0.9998]\\

|\phi_7\rangle=[0.6453, 0.7639]\\|\phi_8\rangle=[0.7724,
0.6351]\\
 \end{array}
 $
 &

$\begin{array}{l}

|\phi_1\rangle=[0.4298, 0.9029]\\|\phi_2\rangle=[0.5133,
0.8582]\\

|\phi_3\rangle=[0.3321, 0.9432]\\|\phi_4\rangle=[0.9703,
0.2419]\\

|\phi_5\rangle=[0.1018, 0.9948]\\|\phi_6\rangle=[0.1415,
0.9899]\\

|\phi_7\rangle=[0.3026, 0.9531]\\|\phi_8\rangle=[0.1908,
0.9816]\\
 \end{array}
 $
 &

$\begin{array}{l} |\phi_1\rangle=[0.0089,
1.0000]\\|\phi_2\rangle=[0.1810, 0.9835]\\

|\phi_3\rangle=[0.6282, 0.7780]\\|\phi_4\rangle=[0.6501,
0.7599]\\

|\phi_5\rangle=[0.4055, 0.9141]\\|\phi_6\rangle=[0.4306,
0.9026]\\

|\phi_7\rangle=[0.1430, 0.9897]\\|\phi_8\rangle=[0.2031,
0.9792]\\
 \end{array}
 $ \\ \hline

$D_7:$&$D_8:$&$D_9:$ \\ \hline

$\begin{array}{l} |\phi_1\rangle=[0.5632,
0.8263]\\|\phi_2\rangle=[0.1064, 0.9943]\\

|\phi_3\rangle=[0.9086, 0.4176]\\|\phi_4\rangle=[0.0576,
0.9983]\\

|\phi_5\rangle=[0.3325,0.9431 ]\\|\phi_6\rangle=[0.0025,
1.0000]\\

|\phi_7\rangle=[0.0900, 0.9960]\\|\phi_8\rangle=[0.7227,
0.6912]\\
 \end{array}
 $
 &
$\begin{array}{l}

|\phi_1\rangle=[0.8502, 0.5265]\\|\phi_2\rangle=[0.1541,
0.9881]\\

|\phi_3\rangle=[0.3624, 0.9320]\\|\phi_4\rangle=[0.3696,
0.9292]\\

|\phi_5\rangle=[0.8279,0.5608 ]\\|\phi_6\rangle=[0.1571,
0.9876]\\

|\phi_7\rangle=[0.4334, 0.9012]\\|\phi_8\rangle=[0.7817,
0.6237]\\
 \end{array}
 $

&

$\begin{array}{l} |\phi_1\rangle=[0.7207,
0.6932]\\|\phi_2\rangle=[0.0995, 0.9950]\\

|\phi_3\rangle=[0.1926, 0.9813]\\|\phi_4\rangle=[0.7677,
0.6408]\\

|\phi_5\rangle=[0.2066,0.9784 ]\\|\phi_6\rangle=[0.0124,
0.9999]\\

|\phi_7\rangle=[0.2994, 0.9541]\\|\phi_8\rangle=[0.84579,
0.5335]\\
 \end{array}
 $ \\ \hline
    \end{tabular}
     \caption{The initial states for
$D_1,D_2,D_3,D_4,D_5,D_6,D_7,D_8,D_9$}
\end{table}
\end{document}